\documentclass{aastex6}
\usepackage{gensymb}
\begin{document}


\title{Exploring The Effects Of Disk Thickness On The Black Hole Reflection Spectrum}



\author{Corbin Taylor\altaffilmark{1,4} and Christopher S. Reynolds\altaffilmark{1,2,3,5}}


\altaffiltext{1}{Department of Astronomy, University of Maryland, 1113 Physical Sciences Complex (Building 415), College Park, MD 20742-2421, USA}
\altaffiltext{2}{Institute of Astronomy, Madingley Road, Cambridge, CB3 0HA}
\altaffiltext{3}{Joint Space Science Institute, University of Maryland, College Park, MD 20742, USA}
\altaffiltext{4}{cjtaylor@astro.umd.edu}
\altaffiltext{5}{csr12@ast.cam.ac.uk}

\begin{abstract}

The relativistically-broadened reflection spectrum, observed in both AGN and X-ray binaries, has proven to be a powerful probe of the properties of black holes and the environments in which they reside. Being emitted from the inner-most regions of the accretion disk, this X-ray spectral component carries with it information not only about the plasma that resides in these extreme conditions, but also the black hole spin, a marker of the formation and accretion history of these objects.  The models currently used to interpret the reflection spectrum are often simplistic, however, approximating the disk as an infinitely thin, optically thick plane of material orbiting in circular Keplerian orbits around the central object. Using a new relativistic ray tracing suite ({\tt Fenrir}) that allows for more complex disk approximations, we examine the effects that disk thickness may have on the reflection spectrum.  Assuming a lamp post corona, we find that finite disk thickness can have a variety of effects on the reflection spectrum, including a truncation of the blue wing (from self-shadowing of the accretion disk) and an enhancement of the red wing (from the irradiation of the central 'eye wall' of the inner disk. We deduce the systematic errors on black hole spin and height that may result from neglecting these effects.

\end{abstract}




\section{Introduction} \label{sec:intro}

Arguably one of the most exotic predictions of General Relativity, black holes are regions of space-time that have causally disconnected themselves from the rest of the Universe due to the complete gravitational collapse of a massive object.  These compact objects are believed to be the engines that drive some of the most energetic phenomena in the Universe \citep{Rees1984}: the accretion of matter on to black holes releases large amounts of energy, producing a multitude of phenomena that have been observed in both the nuclei of active galactic nuclei (AGN) and in X-ray binaries (XRBs).  In the case of AGN, these accretion processes are accompanied by the formation of jets \citep{Begelman+1984} and galactic-scale winds that have appreciable effects on the evolution of their host galaxies.  Despite all of this complex astrophysics, black holes are simple objects, being described by three fundamental parameters: mass ($M$), angular momentum ($J$), and electrical charge.  Due to the efficiency with which excess electrical charge is neutralized by ambient plasma, only the first two parameters are relevant to the study of astrophysical black holes.

The spectral energy distributions (SEDs) of AGN and XRBs are commonly dominated by a thermal component (quasi blackbody emission from the accretion disk, $kT \sim 10$ eV in AGN and $kT \sim 1$ keV in XRBs) and a power-law tail extending to the X-ray band believed to originate from the Compton upscattering of thermal photons by a hot electron corona ($kT \sim 100$ keV).  Superposed on this X-ray power-law continuum are Fe-K$\alpha$ fluorescence lines (with rest-frame energies 6.4-6.97 keV depending upon the charge state of the ion), a complex of blurred emission lines at $\sim 0.3 - 1$ keV (the soft excess), and a broad Compton hump peaking at $\sim 20$ keV.  These features result from the irradiation of optically-thick matter by the hard X-ray continuum and hence are commonly referred to as the reflection spectrum \citep{Basko1978, George+Fabian1991}.  Examining the reflection spectrum, one often finds that the features are strongly broadened, skewed, and redshifted indicating that they originate from the inner-most regions of the accretion disk \citep{Fabian+1989}.

By studying the reflection spectrum, one is able to probe the inner region of the black hole accretion flow (e.g. \citealt{Reynolds+Nowak2003} and references therein). Since the strengths of the relativistic broadening and redshifting effects vary with the location of the inner-most stable circular orbit ($r_{\rm ISCO}$) which, in turn, depends upon the black hole's angular momentum, one is also able to estimate the spin parameter of the black hole ($a \equiv Jc/GM^{2}$) \citep{Iwasawa+1996, Dabrowski+1997, Brenneman+Reynolds2006}.  The measurement of spin is interesting from both an astrophysical and a cosmological perspective.  Examining black hole spin allows us to characterize the effects of frame dragging, a prediction of General Relativity in which a rotating massive object will drag spacetime along with it, and thus has the potential to probe the fundamental nature of gravity. The spin of supermassive black holes also provide a window into their growth histories, delineating accretion-dominated growth from merger-dominated growth \citep{Moderski+1998, Sesana+2014}.  Conveniently, measuring a black hole's spin via its reflection spectrum can be done without knowledge of the mass or distance of the black hole, since all relevant physical quantities can be written in terms of the gravitational radius ($r_{\rm g} \equiv GM/c^{2}$) and thus allowing $M$ to act simply as a scaling constant.

There is a sizable body of work on the use of the black hole reflection spectrum as a probe of black hole physics (see review by \citealt{Reynolds2014} and references therein).  The first such study was by \cite{Fabian+1989} in order to explain the shape of the Fe line observed towards Cyg X-1.  In the same paper, it was suggested that similar analysis could be used to model the soft X-ray excess observed in the Seyfert 1 galaxy MCG-6-30-15 \citep{Nandra+1989,Tanaka+1995}.  These early investigations treated the broad iron line in isolation, but with the deployment of XMM-Newton and subsequent high-throughput, wide-bandpass X-ray observatories (BeppoSAX, Suzaku, and NuSTAR), the improved quality of spectra from AGN and XRBs allows us to consider the full reflection spectrum.

While the specific methodology for characterizing relativistic reflection might differ between software suites, the overall technique involves the calculations of the paths of photons through curved spacetime, tracing the null-geodesics from the disk to the observer at some sufficiently large distance away, with the coronal irradiation of the disk either computed direction via a secondary integration or approximating said irradiation as a power-law as a function of cylindrical radius ($\propto \rho^{-\alpha}$) (\citealt{Dauser+2010} and references therein). One of the most common geometries used in these calculations is of a lamp post corona \citep{Martocchia+Matt1996, Reynolds+Begelman1997, Miniutti+Fabian2004} and an optically thick, razor thin disk.  Strictly, this means that the disk is approximated as a completely opaque, infinitely thin surface (a razor-thin disk) that lies in the plane perpendicular to the black hole's axis of rotation, while the corona is approximated as an X-ray point source of some height $h$ above the black hole on the system's rotational axis.  This point source corona is commonly thought of as a first-order approximation to the base of a polar jet \citep{Biretta+2002,Ghisellini+2004} or gap region of a black hole magnetosphere \citep{Hirotani+Okamoto1998}. It must be emphasized that the exact geometry of AGN coronae is still unknown, and thus the lamp post must be viewed as a fiducial model rather than a rigorous assertion, but recent work using X-ray timing analysis and microlensing suggests that for systems where $L > 10^{-2}\, L_{\rm Edd}$,  the X-ray continuum flux comes from very compact regions (near the pair production limit) that typically are within 3-10 $r_{\rm g}$ from the event horizon and physically separated from the accretion disk (\citealt{Reis+Miller2013}, \citealt{Fabian+2015} and references therein).

Although there has been recent effort to incorporate complex coronal geometries into reflection models by expanding on the lamp post \citep{Wilkins+2016} and comparing these models to observation \citep{Wilkins+2017}, the effects of disk geometry on the reflection spectrum is a scantly explored topic \citep{Pariev+Bromley1998, Wu+Wang2007}.  The recent popularity of the lamppost corona model and the possibility that the disk thickness may be comparable to  coronal heights, demands a renewed exploration of disk thickness on the reflection spectrum. The structure of the disk has been a popular topic of research in modern astrophysics (see the review by \citealt{Blaes2014}), but its exact nature still remains elusive.  Intuitively, one can conclude that a razor-thin disk is unphysical, as the standard pressures (e.g. gas- or radiation-pressure) that arise in the disk as matter accretes onto the black hole would naturally cause the disk to have some non-zero scale height.  With the signal-to-noise data produced using modern astrophysical observatories (e.g. XMM-Newton), one is able to derive constraints on physical parameters, such as black hole spin and coronal height, with relatively small instrumental uncertainties.  As such, an investigation into the systematic error inherent in the assumptions and approximations made in our spectral modeling is necessary.

In order to explore systematic errors in applying these razor-thin disk models to real astrophysical data, we have created a relativistic raytracing suite that allows for the finite thickness disks.  Here we present results using the case of a radiation pressure dominated, optically thick, geometrically thin disk \citep{Shakura+Sunyaev1973} and assuming the validity of the popular lamp post corona model, illustrating the importance of disk geometry in interpreting the black hole reflection spectrum.  In Section \ref{sec: methods}, we describe this new model ({\tt Fenrir}) that allows for non-trivial disk geometries and give a brief description of the primary outputs from the integration (irradiation- and line-profiles). We present the results from {\tt Fenrir} in Section \ref{sec:results} for various values of disk thickness, confirming that there are qualitative differences in the reflection spectrum when the disk has a vertical extent. Using a $\Delta \chi^2$ comparison and standard spectral fitting techniques, we confirm that these qualitative differences translate to significant biases when fitting a contemporary model that neglects disk geometry to data taken from an AGN with modest disk thickness, leading to the underestimation of both black hole spin and coronal height. We end by discussing the possible implications that the predictions from {\tt Fenrir} have on previous attempts to measure AGN properties in Section \ref{sec:discussion}.

\section{Methods} \label{sec: methods}

General relativistic (GR) ray-tracing lies at the heart of calculating the relativistic blurring from a black hole accretion disk.  We have developed a GR ray tracing code that makes use of a fourth-order Runge-Kutta (RK4; \cite{Press+2002}) integrator to calculate the paths of photons through the gravitational potential of a rotating, electrically-neutral black hole, making use of dynamic step refinement in order to properly handle paths that venture close to coordinate singularities.  Such a black hole is described by two fundamental parameters ($M$ and $a$) and has a spacetime is described by the Kerr metric \citep{Kerr1963}.  Setting $r_{\rm g} = 1$ and $c = 1$, one can write this metric in the pseudo-spherical Boyer-Lindquist coordinates \citep{Boyer+Lindquist1967} as

\begin{equation}
\centering
ds^{2} = g_{\rm \mu\nu} dx^{\rm \mu} dx^{\rm \nu} 
= -\left(1-\frac{2r}{\Sigma}\right)dt^{2} - \frac{4ar\ \sin^{2}\theta}{\Sigma}dtd\phi + \frac{\Sigma}{\Delta}dr^{2}+\Sigma d\theta^{2} + \left(r^{2}+a^{2} + \frac{2a^{2}r\ \sin^{2}\theta}{\Sigma}\right)\ \sin^{2}\theta \ d\phi^{2}, 
\label{eq:kerrmetric}
\end{equation}
\newline
where $t$, $r$, $\theta$, and $\phi$ are the four dimensions of spacetime.  $\Delta$ and $\Sigma$ are defined as

\begin{equation}
\centering
\Delta \equiv r^{2} - 2r + a^{2},\newline
\label{eq:kerrdelta}
\end{equation}

\begin{equation}
\centering
\Sigma \equiv r^{2} + a^{2}\cos^{2}\theta.
\label{eq:kerrsigma}
\end{equation}

These are physically important quantities, as the positive root of $\Delta\ = 0$ represents the edge of the event horizon ($r_{\rm +}$) while the value of $r$ at which $\Sigma = 2r$ defines the edge of the ergosphere (i.e. the static limit).  Calculating the Lagrangian [$2\mathcal{L} = (ds/d\lambda)^{2}$] for some affine parameter $\lambda$, one is able to derive the equations of motion for time-like and null geodesics in terms of three conserved quantities: energy ($E$), angular momentum about the axis of rotation ($p_{\phi}$ or $l$), and the Carter constant ($Q$).  See \cite{Bardeen+1972} for a complete description and derivation of these equations of motion.

For this investigation, we use of the lamp post geometry for the irradiating X-ray source, approximating the corona as a stationary X-ray point source that is located on the rotational pole ($\theta = 0$) at some height ($r = h$).  The accretion disk mid-plane is assumed to lie in the $\theta = \pi/2$ and the disk is approximated as a radiatively dominated, optically thick disk with a pressure scale height ($H$) of \cite{Shakura+Sunyaev1973}.

\begin{equation}
\centering
H = \frac{3}{2}\frac{1}{\eta}\left(\frac{\dot{M}}{\dot{M}_{\rm Edd}}\right)\left[1 - \left(\frac{r_{\rm ISCO}}{\rho}\right)^{\frac{1}{2}}\right]
\label{eq:ss73}
\end{equation}
where $\rho = r\sin \theta$ is the pseudo cylindrical radius, $\eta$ is the radiative efficiency of the disk, and $r_{\rm ISCO}$ is the radius of the inner-most stable circular orbit, assumed to be the disk's inner-edge.  As the exact structures of AGN accretion disks is a topic still being studied, we chose \ref{eq:ss73} as a simple fiducial model from which to begin to incorporate accretion disk geometry into the raytracing paradigm. While this model is strictly a Newtonian solution, the profile of $z$/$\rho$ is qualitatively similar to that of a fully relativistic thin accretion disk \citep{Novikov+Thorne1973}, only differing in their normalizations. As such, we expect no qualitative difference in our results by making this simplifying assumption.

\begin{figure*}
\centering
\includegraphics[width=0.96\linewidth]{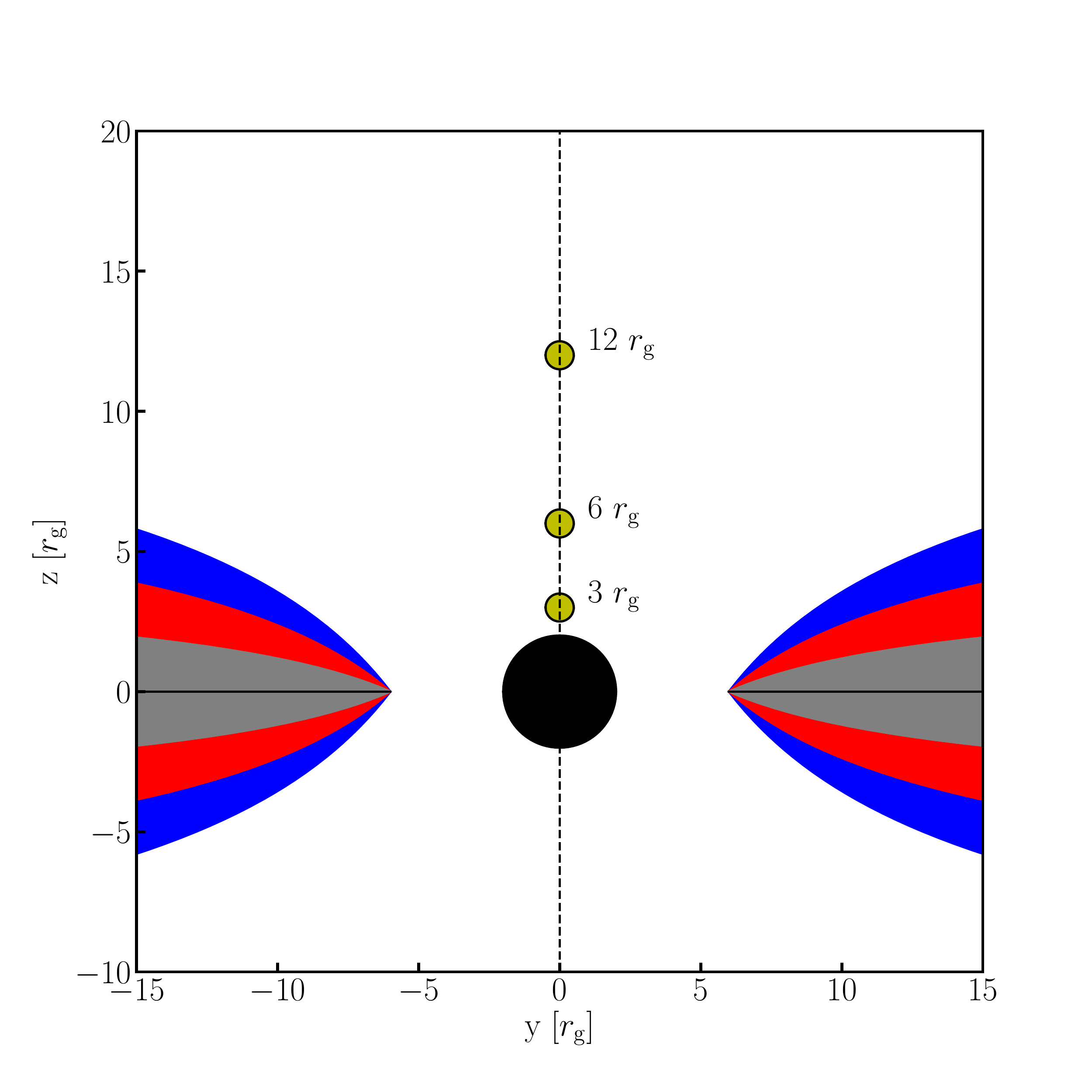}
\caption{An illustration of a Schwarzschild black hole (central black circle), an accretion disk, and lamppost coronae (yellow circles) at various heights ($h$ = 3 $r_{\rm g}$, 6 $r_{\rm g}$, and 12 $r_{\rm g}$) along the rotation axis (dashed line). The disk profile is shown for four different values of the Eddington fraction ($\dot{M}$/$\dot{M}_{\rm Edd}$): 0.0 (black, equivalent to the zero-thickness approximation), 0.1 (grey), 0.2 (red), and 0.3 (blue). The disk half-thickness is given by $z$ = 2$H$ where $H$ is the radiative pressure scale height of a radiatively dominated, optically thick, geometrically thin disk \citep{Shakura+Sunyaev1973} given in Equation \ref{eq:ss73}. Thermal photons from the disk are Compton upscattered into the X-ray regime by the hot electron corona, which then either propagate to infinity (producing the Compton X-ray power-law observed in AGN) or irradiate the disk and are reprocessed (producing the reflection spectrum). As one can see, as the accretion rate increases, the disk thickness becomes comparable to that of the coronal height, suggesting that the geometry could have substantial effects on the irradiation of the disk and the resulting reflection spectrum.}
\label{cartoon}
\end{figure*}

\begin{figure*}
\centering
\includegraphics[width=\linewidth]{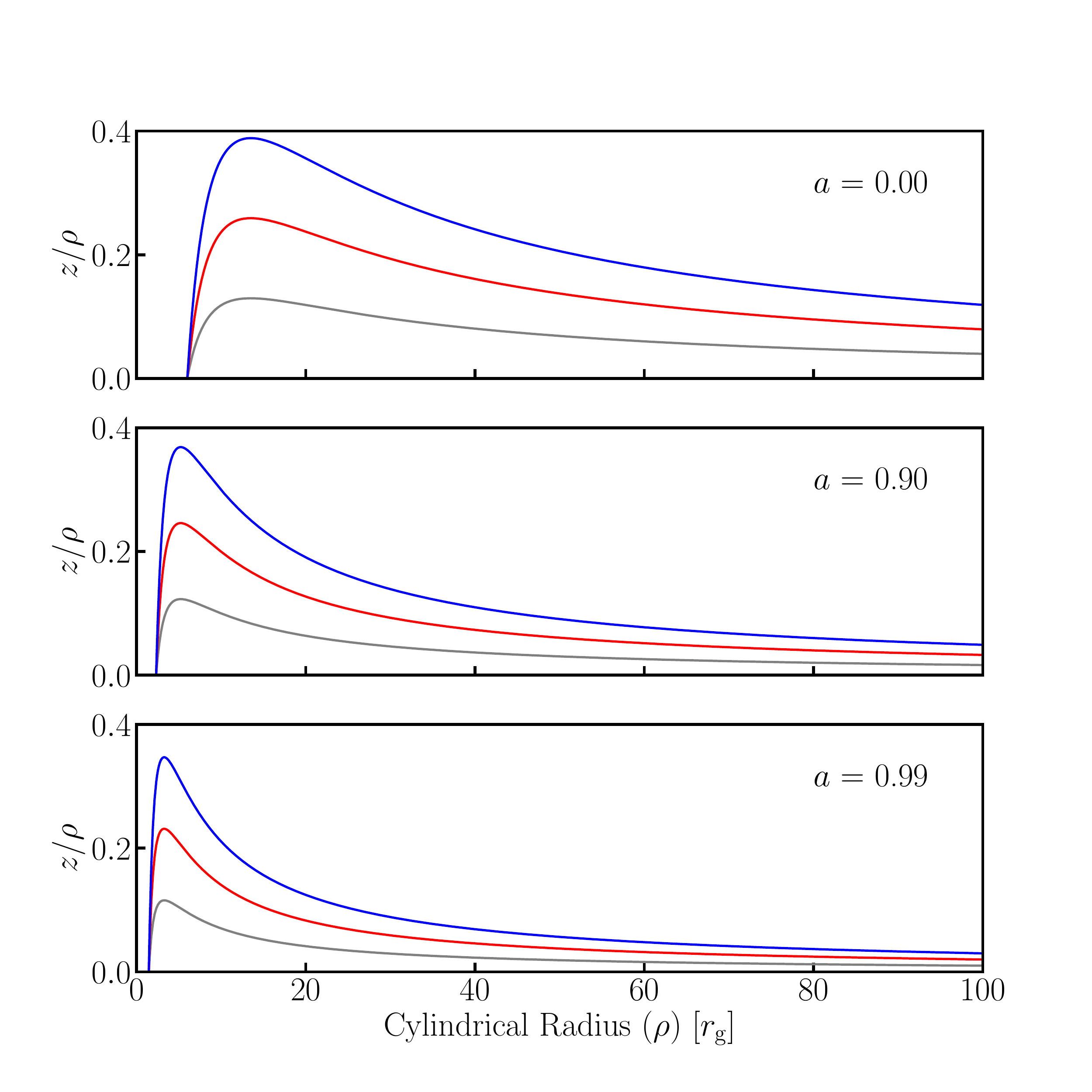}
\caption{The geometric thickness profiles of the disk, where $z$ = 2$H$ ($H$ being defined by Equation \ref{eq:ss73}) and $\rho$ is the pseudo-spherical radius ($\rho$ = $r$sin $\theta$). We show $z$/$\rho$ for three different spin values ($a$ $\in \{0.00,\ 0.90,\ 0.99\}$, denoted in upper right of panels) and three different Eddington accretion ratios: $\dot{M}$/$\dot{M}_{\rm Edd}$ = 0.1 (gray), 0.2 (red), 0.3 (blue). The inner edge of the disk is assumed to be at the inner-most stable circular orbit $\rho$ = $r_{\rm ISCO}$ and the disk has no net torque at its inner edge. For any value of $a$, $z$/$\rho$ increases $\dot{M}$, while for any given accretion rate, $z$/$\rho$ decreases with increasing $a$.}
\label{thickness_panel}
\end{figure*}

In this work, we do not consider reflection signatures from material within the inner-most stable circular orbit, appropriate if the production of X-rays is efficient thereby strongly ionizing this region \citep{Reynolds+Begelman1997}. Both $\eta$ and $r_{\rm ISCO}$ are functions of $a$, with the efficiency increasing with black hole angular momentum and $r_{\rm ISCO}$ decreasing with $a$, resulting in the disk becoming thinner with increasing spin for a given Eddington ratio.  In the context of the study, the Eddington ratio ($\dot{M}$/$\dot{M}_{\rm Edd}$) acts as a disk thickness parameter, which for a given value of spin, increases the geometric thickness of the disk as it itself increases.

When considering the effects of finite disk thickness on the X-ray reflection spectrum, one is really concerned with the height of the X-ray scattering photosphere. This has a non-trivial connection with the pressure scale-height requiring a full radiative transfer calculation to handle properly. Here, as a first exploration of the effects of disk thickness, we simply assume that the X-ray photosphere is located at a height of $z = 2H$. In Figure \ref{cartoon} we present an illustration of such a disk around a central Schwarzschild black hole along with the lamppost corona at various heights, showing that for some values of $\dot{M}$, the thickness of the disk becomes comparable to $h$. Figure \ref{thickness_panel} shows the ratio of disk height and cylindrical radius ($z$/$\rho$) for three different values of spin ($a$ = 0.00, 0.90, 0.99), showing how as $a$ increases, the inner edge of the disk ($r_{\rm ISCO}$) grows closer to the central black hole. Also, it is clear by this Figure that as $\dot{M}$ increases for a given spin, $z$/$\rho$ increases as expected from Equation \ref{eq:ss73}.

We assume that the disk rotates cylindrically: all matter in the disk at a pseudo cylindrical radius $\rho$ is assumed to have an orbital three-velocity ($V^{\phi}$) equal to that of the material at the same cylindrical radius in the orbital mid-plane ($\theta = \pi/2$) orbiting in prograde circular Keplerian orbits.. This mid-plane orbital velocity is taken from \cite{Bardeen+1972} and is equal to $V^{\phi} = (r^{\frac{3}{2}} + a)^{-1}$, while the disk elements' four-velocity vectors (\textbf{\textit{U}}) are calculated as to be consistent with Lorentz invariance (i.e. $\textbf{\textit{U}} \cdot \textbf{\textit{U}} = -1$).

We trace photon trajectories from the lamppost corona down to the photosphere of the disk, and then from the disk to a stationary observer at some inclination angle $i$ and a radial distance of $r=10^{3}\,r_{\rm g}$.  The former trajectory is traced forward in time, integrating each photon individually and assuming that the photons are emitted isotropically in the corona's rest frame \citep{Karas+1992, Reynolds+1999}.  For computational efficiency, we have traced from the disk to the observer backwards in time, starting at a point on the observer's image plane and ending at its point of intersection with the disk.  This eliminates the need to calculate the paths of photons that do not fall within the observer's line of sight, and thus do not contribute to the observed spectrum.  There are two main outputs of the raytracing, the irradiation profile across the disk and the disk image, which when integrated together give the (un-normalized) profile of a $\delta$-function emission line that has been skewed and broadened (i.e. blurring kernel).  We will denote the irradiation profile as $\mathcal{E}(\rho , \phi)$, and in this case $\mathcal{E}(\rho , \phi)$ = $\mathcal{E}(\rho)$ due to the axisymmetry of the disk geometry.  

In detail, the energy of the incident photon from the corona on the disk element at ($r$, $\theta$, $\phi$) is calculated via the dot product of the photon's 4-momentum ($p^{\nu}$) and the disk element's 4-velocity ($U^{\mu}$).  The photons are binned cylindrically into annuli along the surface of the disk that have a co-moving surface area of

\begin{equation}
\centering
A_{\rm ann} = 2\pi \gamma \bigg[g_{\rm rr} + g_{\rm \theta \theta}\left(\frac{d\theta}{d\rho}\right)^{2} \sin^{2}\theta\bigg]^\frac{1}{2} \left(\frac{g_{\rm \phi \phi}}{\sin^{2}\theta}\right)^{\frac{1}{2}}d\rho
\label{eq:annulus}
\end{equation}
where $\gamma$ is the Lorentz factor of the disk flow calculated in a locally non-rotating reference frame constructed at the same point in spacetime as the disk element.  This is a generalized equation from that which is presented in \cite{Wilkins+Fabian2012}, where the non-zero disk thickness is captured by the $d\theta / d\rho$ term. Defining $E_{\rm cor}$ to be the energy of the photon in the coronal rest frame and $E_{\rm disk}$ to be the energy of the same photon as seen by an observer co-moving with the disk, one can find the total photon flux incident on each annuli

\begin{equation}
\centering
\mathcal{E}(\rho) \propto  \sum_{\rm photons} \frac{(E_{\rm disk}/E_{\rm cor})^{\Gamma}}{A_{\rm ann}(\rho)}.
\label{eq:emmisivity}
\end{equation}
where $\Gamma$ is the photon index of the coronal photon distribution ($N_{E} \propto E^{-\Gamma}$). This expression is equivalent to that of Equation 18 of \cite{Dauser+2013}, but we have collapsed the Lorentz factor ($\gamma$) into the area of the annulus given in Equation \ref{eq:annulus}. The additional $\sin \delta$ factor quoted by the authors is not needed in our calculation due to the technical differences between our model calculation and their own. This is the irradiation profile, and for this work, we assume that the emissivity profile of the disk is directly proportional to the irradiation profile.

The second output of the ray-tracing is the map from the disk surface to the image plane of the distant observer. The energy of the observed photon is denoted as $E_{\rm obs}$ and we will define $g \equiv E_{\rm obs}/E_{\rm em}$, where $E_{\rm em}$ is the energy of the photon when it is emitted, as seen from an observer co-moving with the emitting disk element.  The line profile [$\Phi(g)$], which is the specific photon number flux, is calculated as

\begin{equation}
\centering
\Phi(g) \propto \int_{\rho, \rm disk}\int_{g'} \mathcal{E}(\rho) g'^{3}\delta(g - g')\,dg'\,d\rho
\label{eq:lineprofile}
\end{equation}
where the factor of $g^{3}$ in the integrand comes from Liouville's Theorem: specific energy intensity goes as $I_{E} \propto E^{4} \propto g^{4}$ and thus the specific number intensity $\propto g^{3}$.

\section{Results}\label{sec:results}

Using the methods described in Section \ref{sec: methods}, we calculate irradiation profiles, disk images, and line profiles for various values of the spin parameter ($a$), viewing inclination ($i$), corona height ($h$), and mass accretion rate ($\dot{M}$).  In this investigation on the effects of disk thickness on black hole reflection spectra, we have chosen to sample the parameter space coarsely but broadly.  We have chosen to focus on a set of three values each for $a$, $i$, and $h$ ($a \in \{0.00, 0.90, 0.99\}$, $i \in \{15\degree, 30\degree, 60\degree\}$, and $h \in \{3 r_{\rm g}, 6 r_{\rm g}, 12 r_{\rm g}\}$), while for $\dot{M}$ we have chosen to focus on $\dot{M}$/$\dot{M}_{\rm Edd}$ $\in \{0.0,\ 0.1,\ 0.2,\ 0.3\}$. Recall that $\dot{M}$/$\dot{M}_{\rm Edd}$ is simply our proxy for disk thickness. In the $\dot{M}=0.0\ \dot{M}_{\rm Edd}$ case, our model reduces to the standard razor-thin case and hence, can be directly compared with existing models. As part of our code validation, we have confirmed that our model reproduced the {\tt RELXILL} line profiles at this limit \citep{Garcia+2014,Dauser+2014}, as well as the line profiles from the {\tt ky} models \citep{Dovciak+2004}. The maximum accretion rate considered $\dot{M}=0.3\ \dot{M}_{\rm Edd}$ is set by approximate limits on the validity of the thin-disk geometry implicit in Eq. \ref{eq:ss73}.

\subsection{Irradiation Profiles and Disk Images}\label{sec:disks}

\begin{figure*}
\centering
\includegraphics[width=\linewidth]{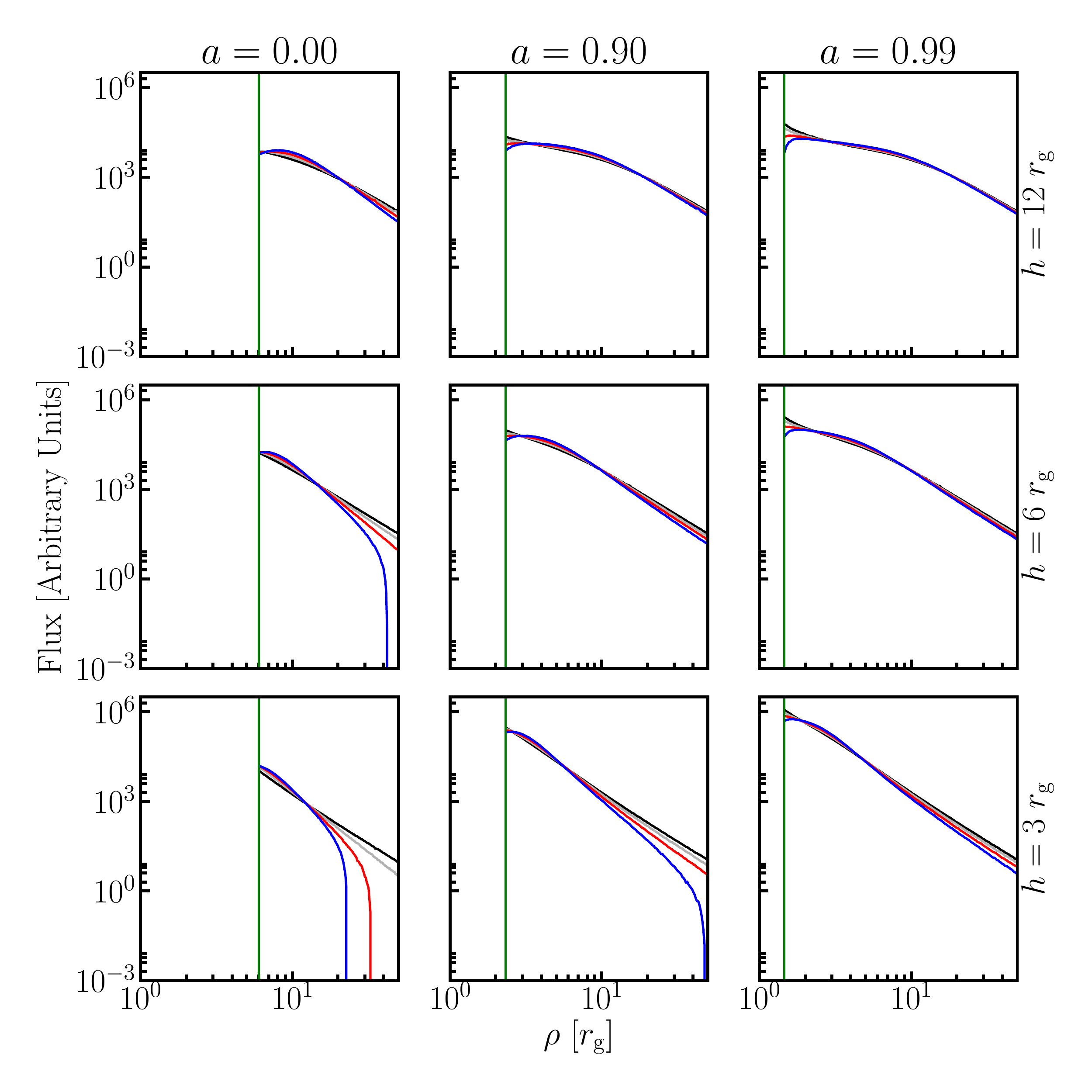}
\caption{The irradiation profiles $\mathcal{E}(\rho)$ for various spins ($a$, columns) and lamp post coronal heights ($h$, rows). The different colors represent different levels of accretion: $\dot{M}/\dot{M}_{\rm Edd} = $ 0.0 (black), 0.1 (grey), 0.2 (red), and 0.3 (blue). The horizontal axis of each panel represents the cylindrical radius ($\rho$), while the vertical axis represents the photon flux incident on the disk at that particular radius.  Finally, the vertical green line represents the $\rho$ that corresponds to the inner-most stable circular orbit of the disk ($r_{\rm ISCO}$), which varies with $a$. For the case of $a = 0.00$, one finds an enhancement of the irradiation profiles for smaller radii, while there appears to be a notable decrease in the incident flux further out on the disk (the flux becomes negligible for $\rho > 20 r_{\rm g}$ in the most extreme case). A similar trend occurs for $a = 0.90$ and 0.99, but not as dramatically, consistent with a thinner disk due to a larger efficiency factor ($\eta$) at these higher spin values (see Equation \ref{eq:ss73}). The profiles for these cases flatten near $r_{\rm ISCO}$, and turn over slightly near $r_{\rm ISCO}$. Overall, we can conclude that the predicted lamppost irradiation profile is heavily dependent on the geometry of the reflection surface.}
\label{emProfiles}
\end{figure*}

We present the resulting irradiation profiles in Figure \ref{emProfiles}.  In the Schwarzschild case ($a = 0$), for all three corona heights presented, one finds that increasing the disk thickness (i.e. increasing the black hole accretion rate) results in an increase in the incident flux at small cylindrical radii ($< 20\ r_{\rm g}$) and less illumination at larger values of $\rho$.  These effects are clearly dependent on $h$, especially the drop off at large radii.  The irradiation is prominent for $> 50\ r_{\rm g}$ in the case of $h = 12\ r_{\rm g}$, while it becomes negligible beyond $\sim 40\ r_{\rm g}$ for $h = 6\ r_{\rm g}$ and $\sim 20\ r_{\rm g}$ for $h = 3\ r_{\rm g}$.  The drop off in $\mathcal{E}$ is explained via disk shadowing, where the disk thickness naturally allows for the inner regions of the disk to shield the outter regions from coronal photons.  As will be discussed later, shadowing will have a prominent effect on the intensity of the reprocessed light across the surface of the disk, and hence the line profile.

A similar trend is seen for the higher spin cases, with the effects becoming less prominent at larger values of spin.  Physically, this is due to the decrease in thickness of the accretion disk with increasing spin, since $\eta$ increases with $a$, and efficiency is inversely proportional to disk scale height (see Eq \ref{eq:ss73}).  Interestingly, at the inner-most radii, the irradiation profiles actually start to decrease with $\dot{M}$, flattening and, in certain cases, slightly turning over. This is due to the convex geometry of the inner disk resulting in smaller effective area for the annuli near $r_{\rm ISCO}$, leading to less total photons irradiating the central-most regions of the accretion disk, as well as the annuli at slightly larger values of $\rho$ being able to better intercept photons due to their respective annuli being in closer proximity to the irradiating corona compared to the case of a razor-thin disk.

\begin{figure*}
\centering
\includegraphics[width=0.48\linewidth]{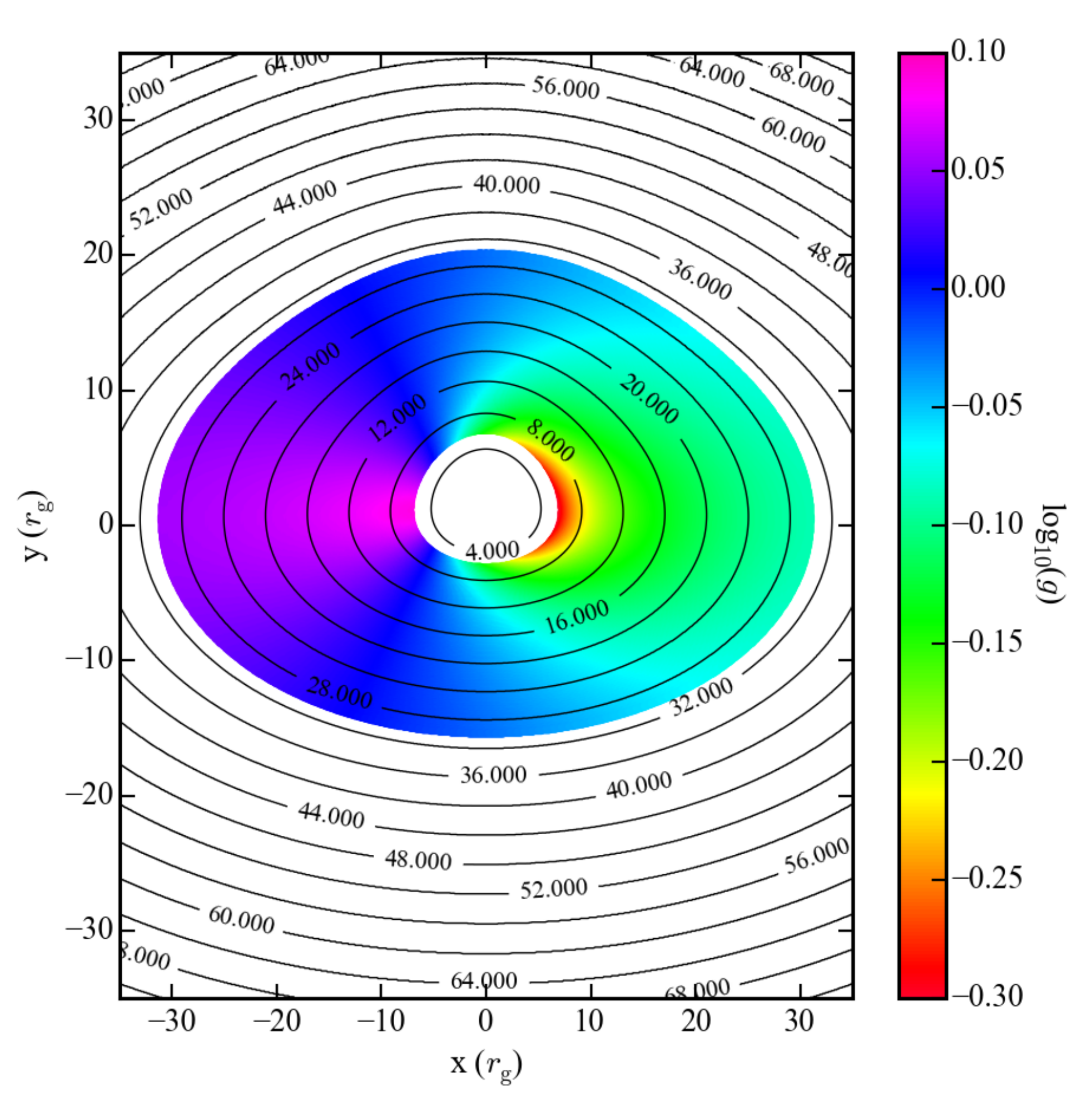}
\includegraphics[width=0.48\linewidth]{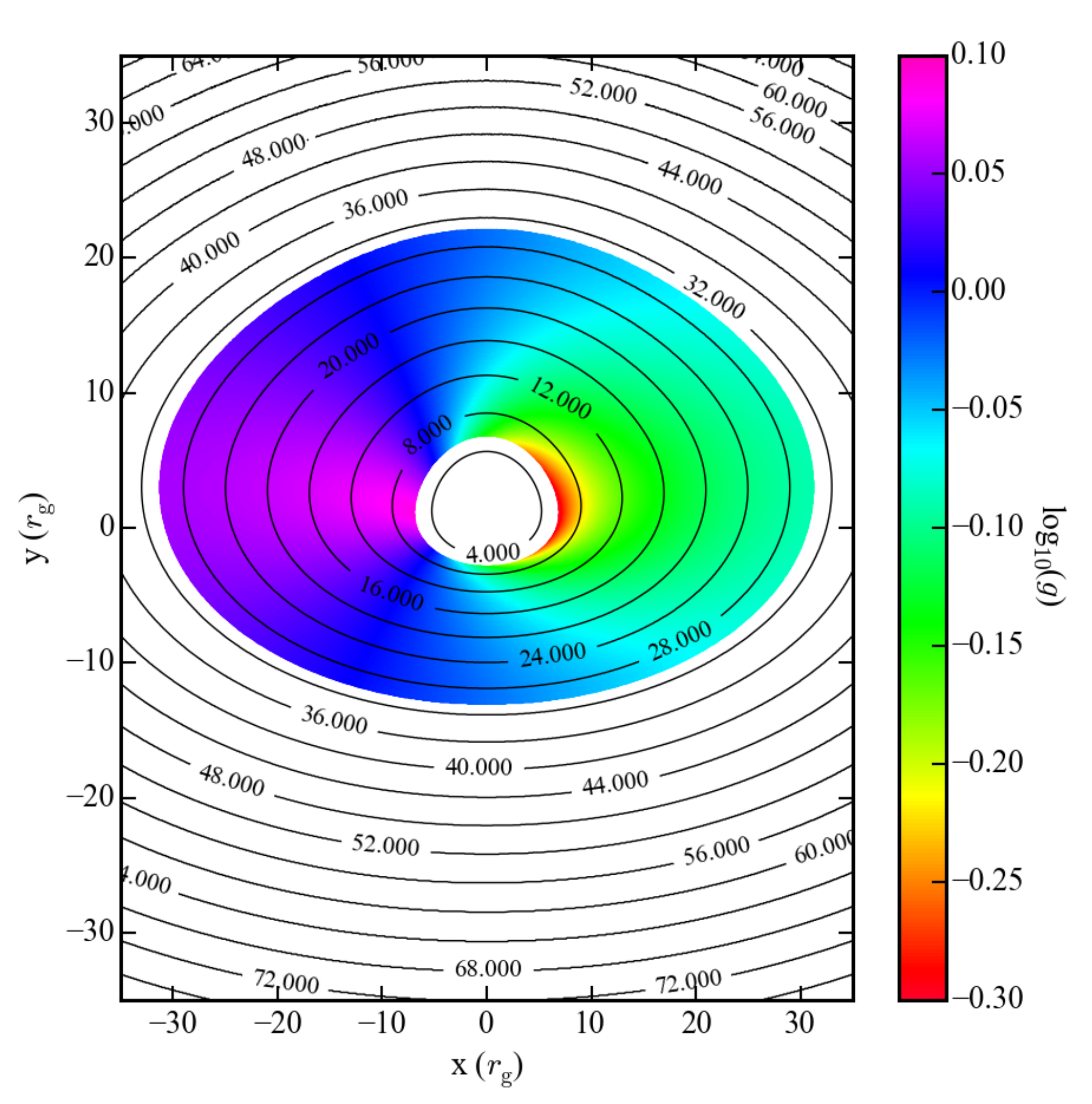}
\includegraphics[width=0.48\linewidth]{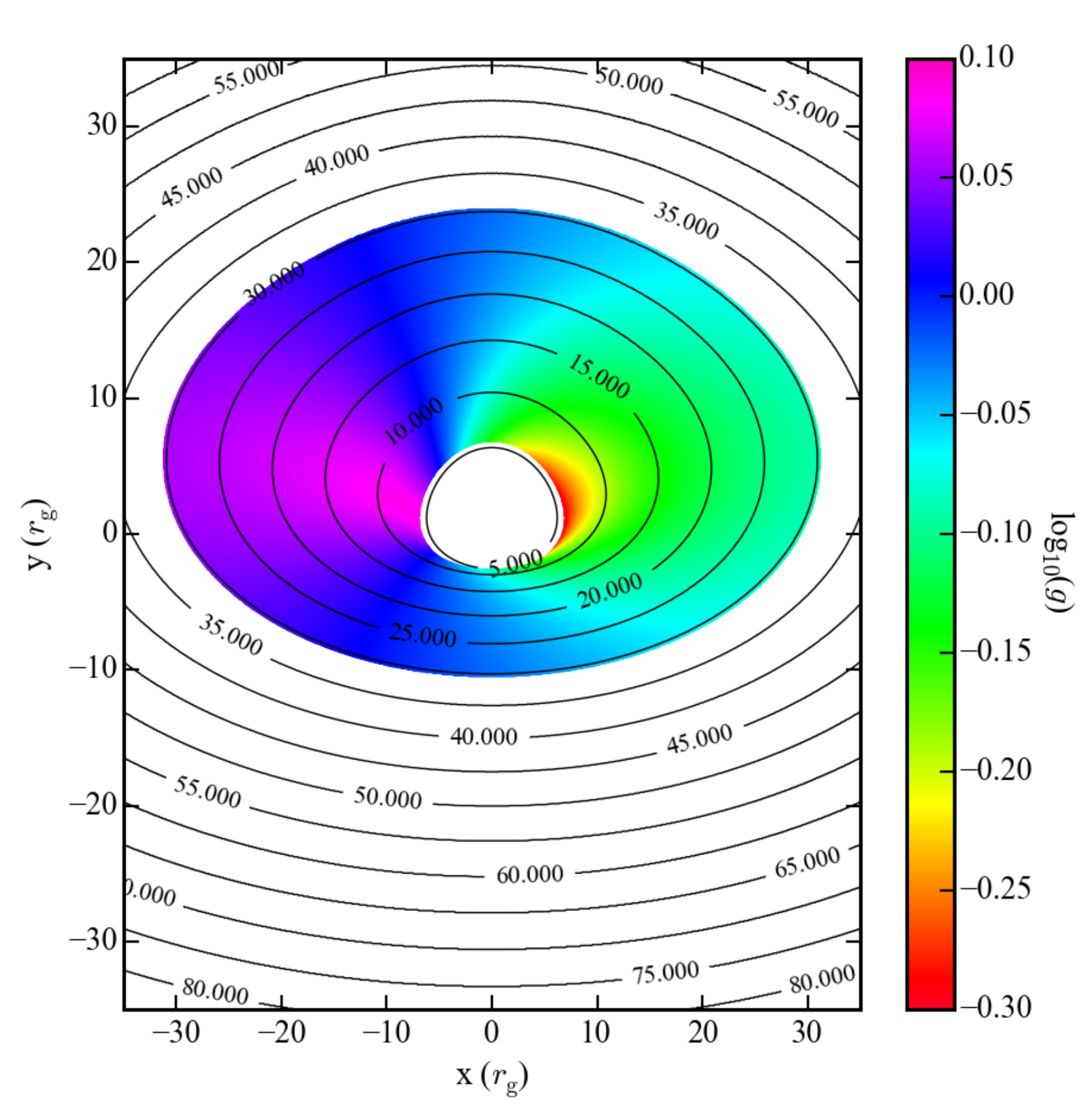}
\includegraphics[width=0.48\linewidth]{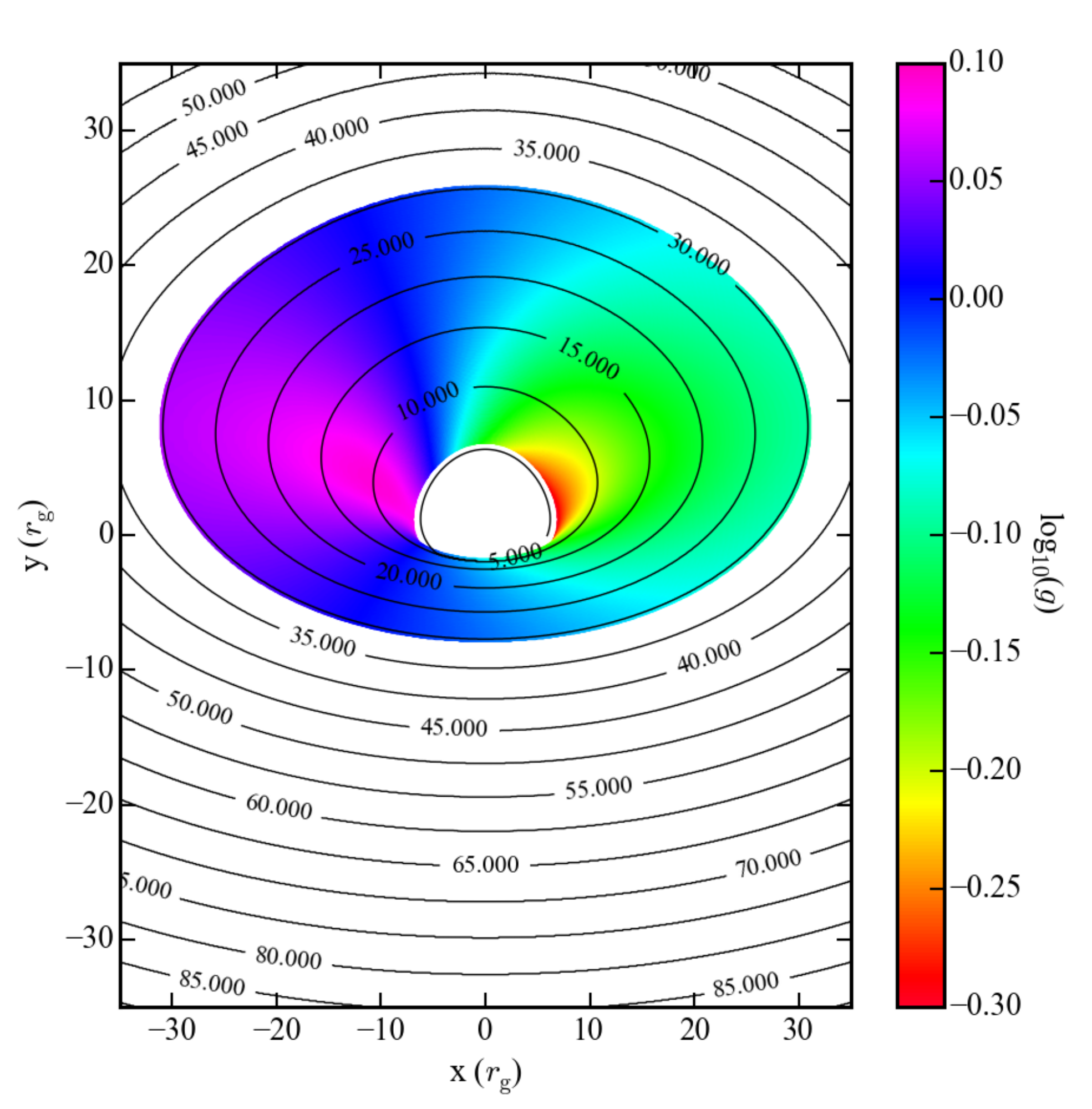}
\caption{Photon energy shifts mapped on to the image plane produced by tracing photons from the disk to a stationary observer at $r = 10^{3}$ $r_{\rm g}$ and an inclination angle of $i = 60\degree$ relative to the axis of rotation.  Each panel represents a different level of disk thickness, corresponding to different levels of accretion: $\dot{M}$/$\dot{M}_{\rm Edd}$ = 0.0 (upper left), 0.1 (upper right), 0.2 (lower left), and 0.3 (lower right). The color scale is log$_{10}$($g$), where $g$ represents the ratio between the energy of the photon as it is emitted by the disk, in the disk's frame, and the energy as observed by the distant observer (i.e. $E_{obs}$ = $gE_{em}$). The contours are that of cylindrical radius ($\rho$).  As one can see, as the thickness of the disk increases, there is a notable change in the geometry of the disk, the photosphere transforming from a flat plane to a bowl shape as one would expect from a Shakura \& Sunyaev (1973) accretion disk. One also sees a slight increase in the subtended angle of the strongly blue-shifted region of the disk.}
\label{gDisks}
\end{figure*}

A sample of redshift maps are presented in Figure \ref{gDisks} for various values of $\dot{M}$, focusing on the Schwarzschild case with an inclination angle of $60\degree$.  The image has been oriented such that the material located on the left side of the image is coming towards the observer, while the right side is receding.  While in principle our models allow for any possible value of the outermost cylindrical radius of the disk ($R_{\rm out}$), we have chosen $R_{\rm out} = 30\,r_{\rm g}$ for convenience. One can see that, as the accretion rate increases, the geometry of the disk changes accordingly, going from a flat plane ($\dot{M} = 0.0\ \dot{M}_{\rm Edd}$ in the upper left) to more of a funnel ($\dot{M} = 0.3\ \dot{M}_{\rm Edd}$, lower right).  There also appears to be a small increase in the subtended solid angle of the region that is emitting the most blueshifted photons, located on the left side of the disk.  As we will see later, this effect, along with the enhancement of disk irradiation at the corresponding $\rho$ will translate to an enhancement of the blue peak of the line profile when $a = 0$.

\begin{figure*}
\centering
\includegraphics[width=0.48\linewidth]{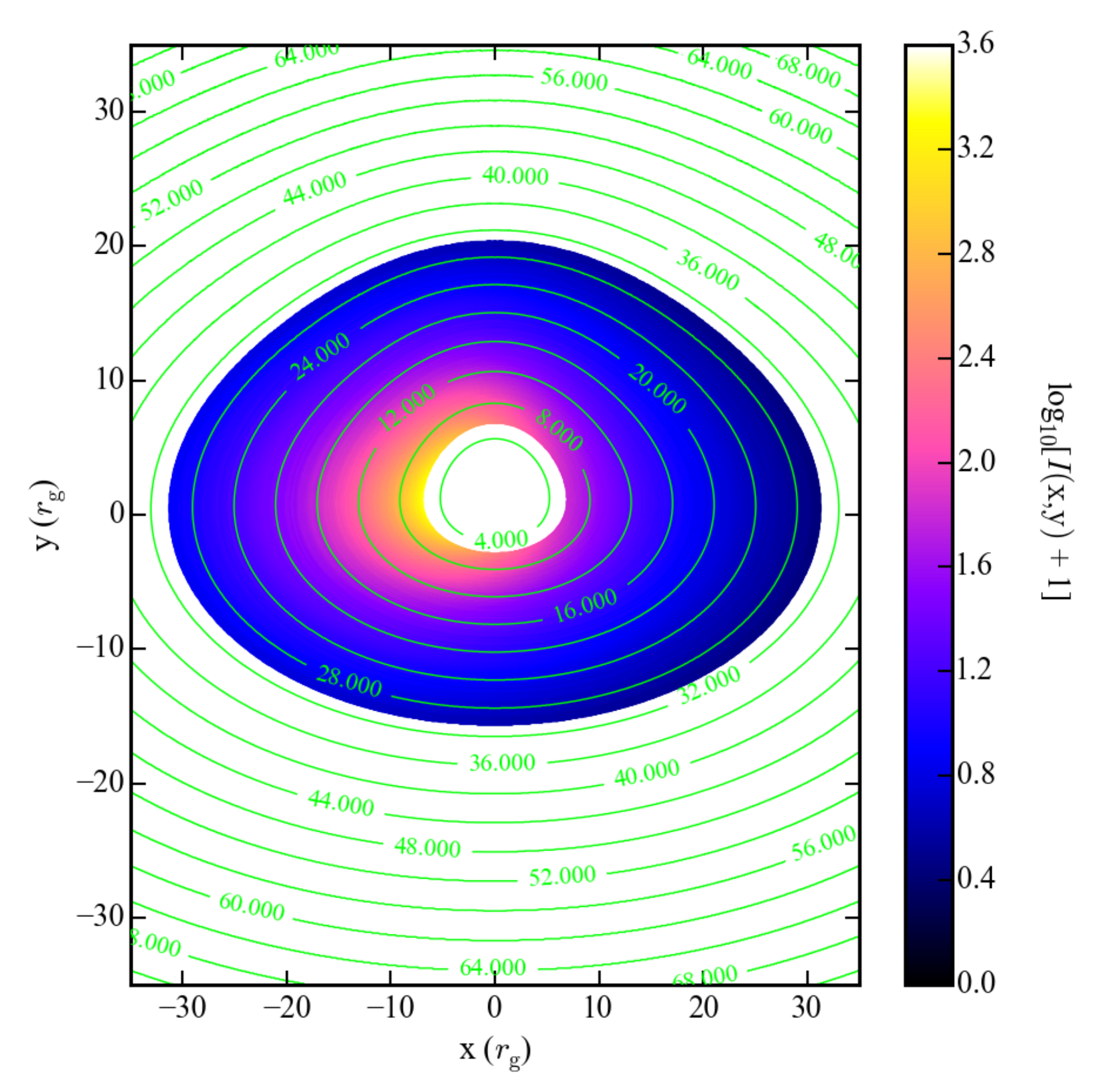}
\includegraphics[width=0.48\linewidth]{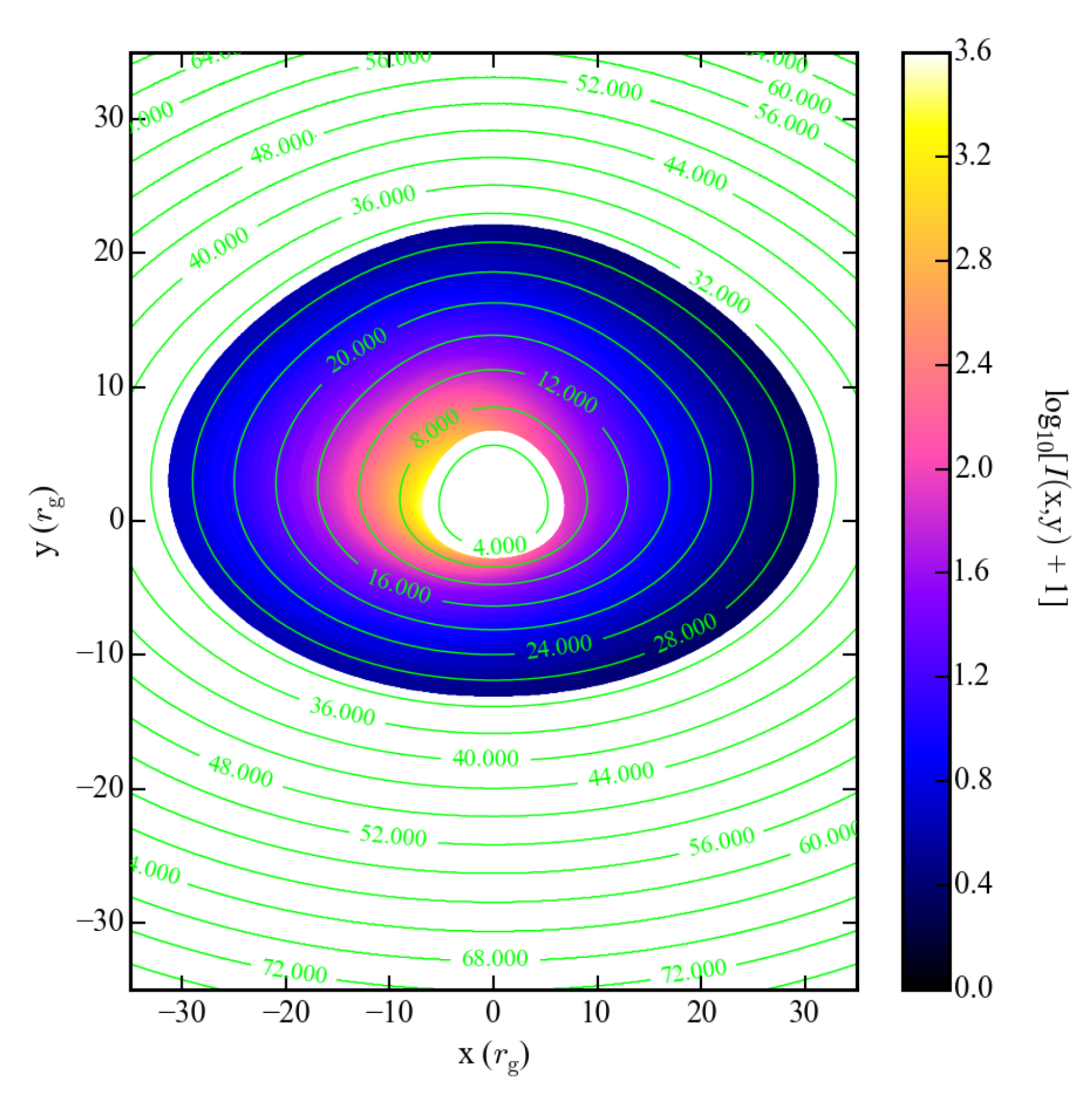}
\includegraphics[width=0.48\linewidth]{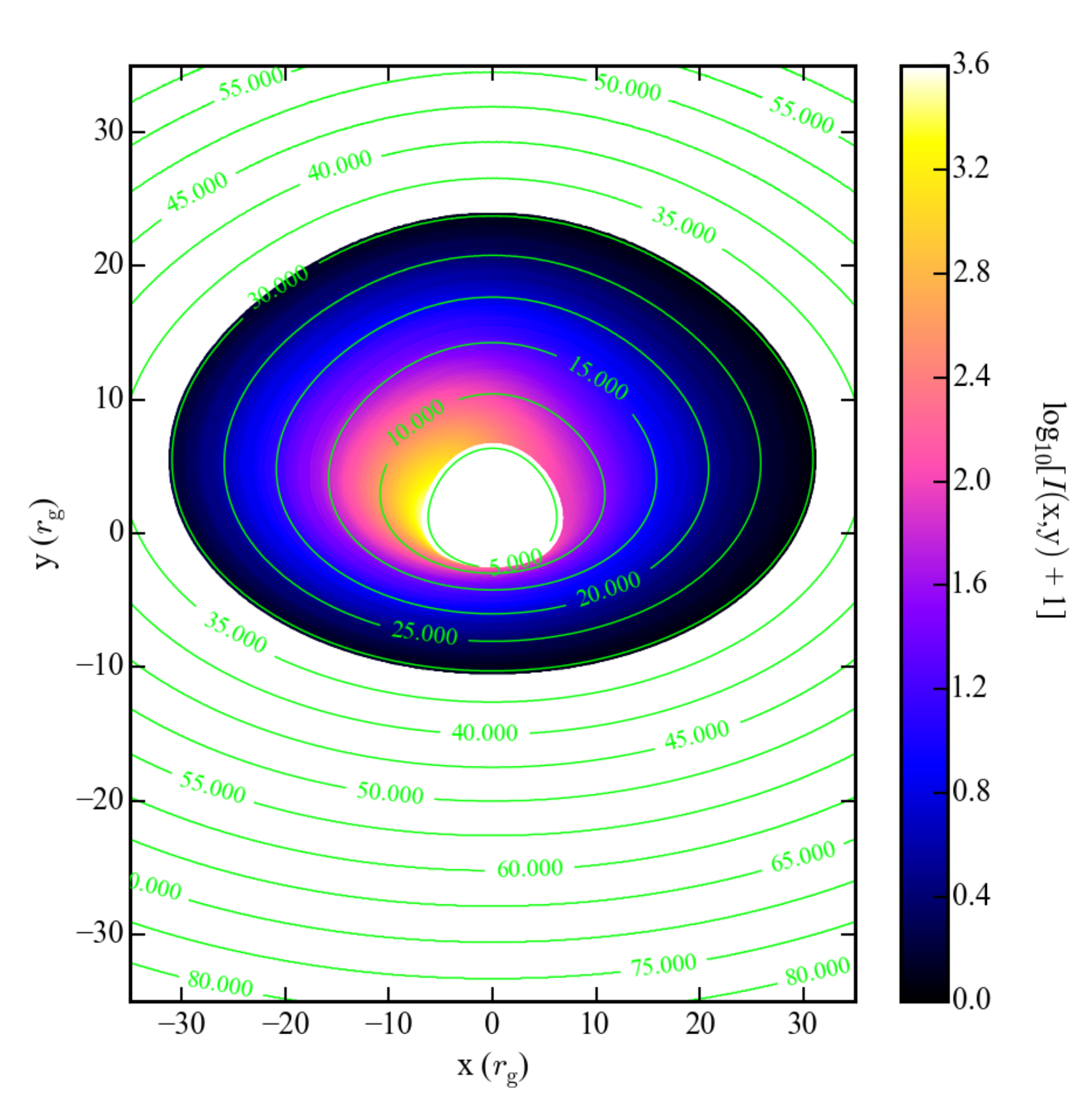}
\includegraphics[width=0.48\linewidth]{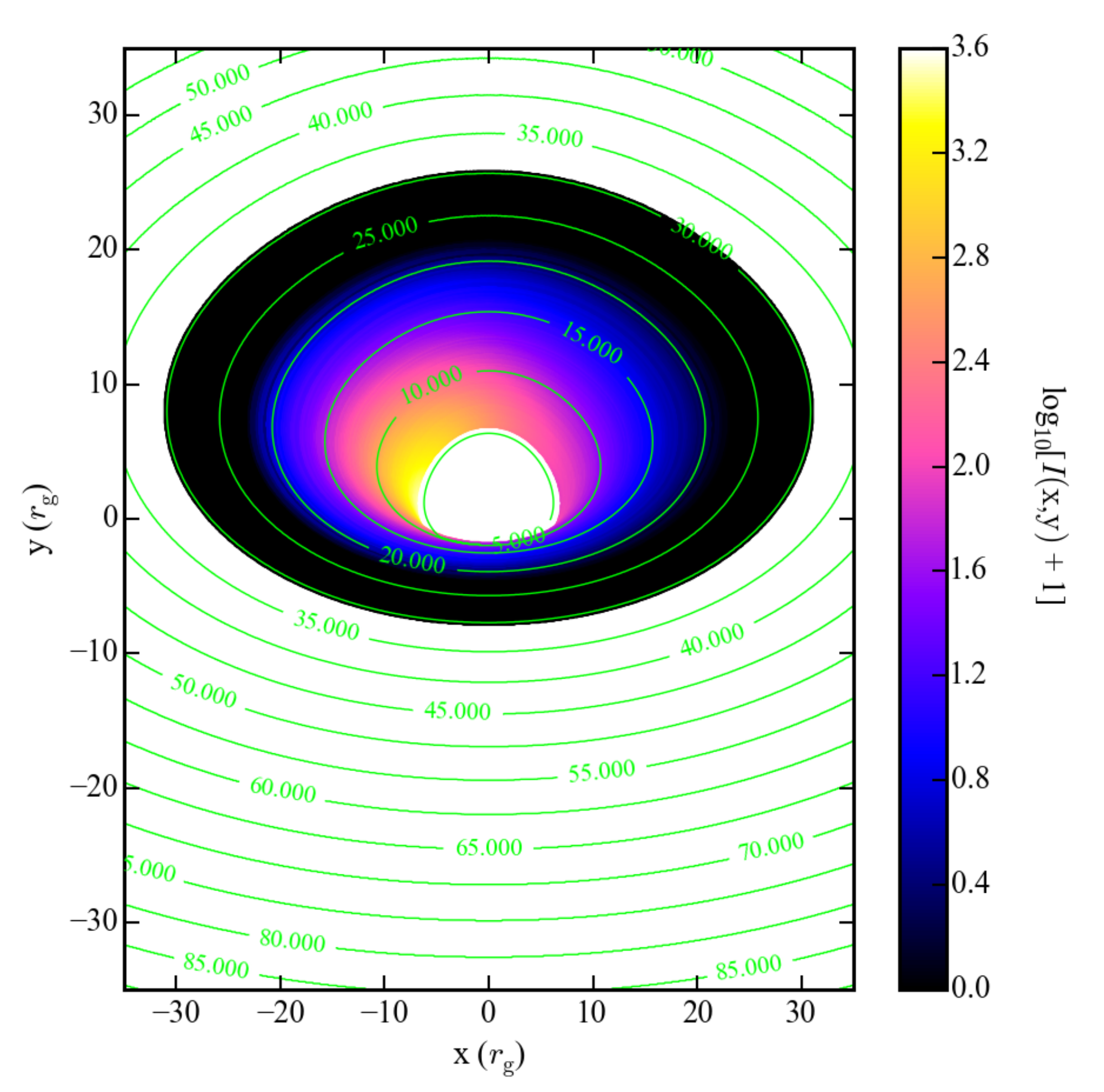}
\caption{Energy intensity maps [$\propto \ \mathcal{E}(\rho)g^{4}$] for the cases that are presented in Figure \ref{gDisks} ; the added value of unity is to correct for null values that occur within the logarithm.  The reflection spectrum intensity is found via combining the disk images with the irradiation profile from a corona of $h = 3$ $r_{\rm g}$.  We see that as $\dot{M}$ increases, the inner regions of the disk act to shield the outer regions from corona radiation, resulting in a suppression of $I$ at larger cylindrical radii.}
\label{iDisks}
\end{figure*}

Finally, we present the intensity map of the reflection spectrum as a function of (x,y) on the image plane [$I$(x,y)] in Figure \ref{iDisks}, using the same values for $a$ and $i$ as Figure \ref{gDisks} and a corona height of $h = 3 r_{\rm g}$.  As suggested by Figure \ref{emProfiles}, the inner regions of the disk act to obscure the material at larger radii from the corona's X-ray photons, and thus these outer regions contribute less to the reprocessing of the coronal photons and the resulting reflection spectrum.  This results in a drastic drop in intensity at larger radii, having a steeper gradient with increasing disk thicknesses.  This set of examples strongly suggests that disk geometry may have a dominant effect reflection spectrum and underscores the motivation for our study.

\subsection{Line Profiles} \label{sec:lines}

\begin{figure*}
\centering
\includegraphics[width=\linewidth]{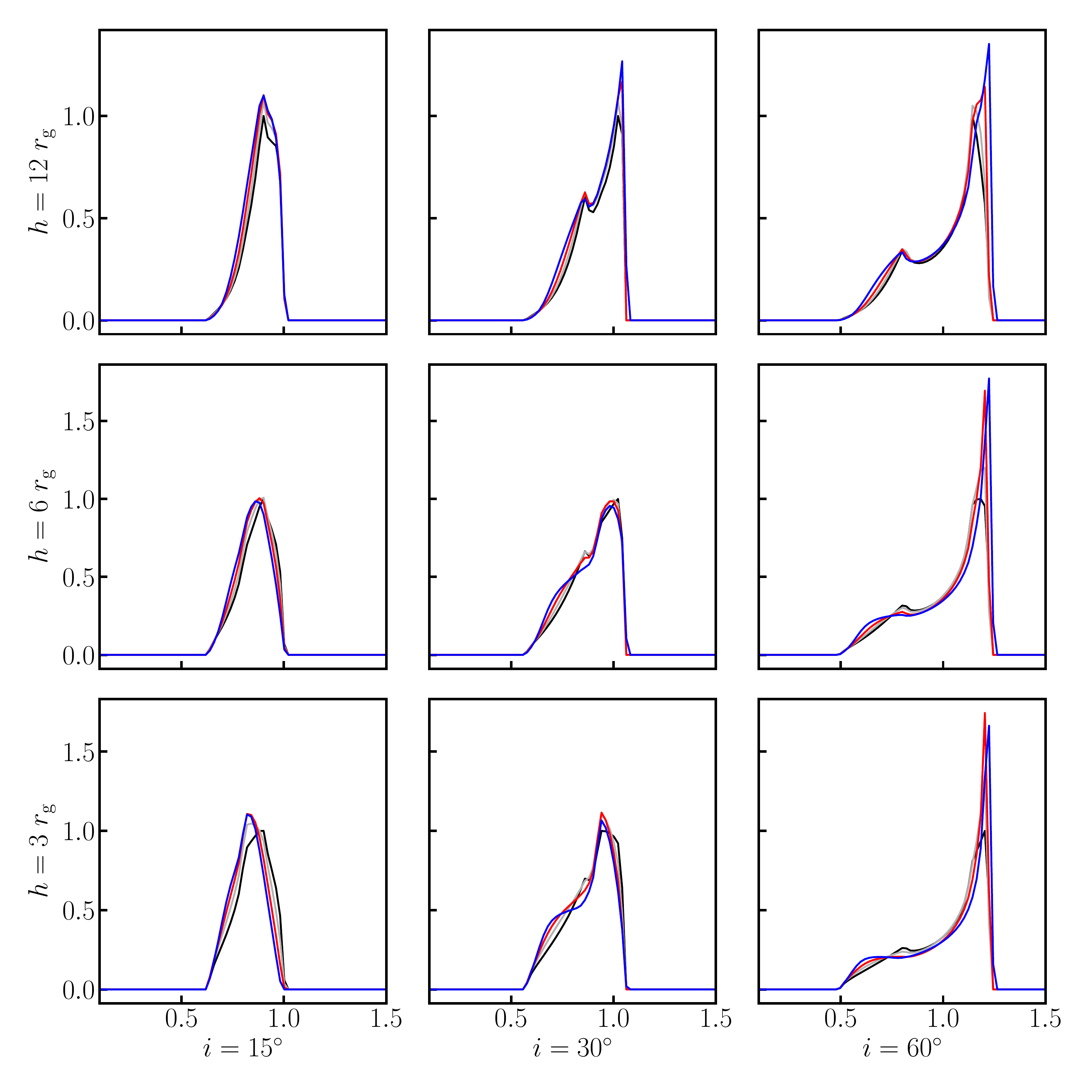}
\caption{A collection of line profiles ($\Phi(g)$) for a Schwarzschild black hole ($a = 0$). The horizontal rows scan through different coronal heights $h$ in units of r$_{\rm g}$ and the vertical columns represent different values of the disk inclination angle $i$, where a lower inclination represents a disk that is more face-on than a higher inclination disk where more of the disk edge is observed.  The horizontal axes are in terms of $g \equiv E_{\rm obs}/E_{\rm em}$ and the vertical axes are photon flux per energy bin, normalized so that the peak of the razor thin disk ($\dot{M} / \dot{M}_{\rm Edd} = 0$) is equal to unit.  The colored lines represent the different accretion rates (or disk thicknesses), with $\dot{M} / \dot{M}_{\rm Edd} = 0$ being black, 0.1 being grey, 0.2 being red, and 0.3 being blue.  One can see that the effect of disk thickness on the line profile is a non-trivial function of disk half-thickness ($z$), $h$, and $i$.}
\label{trans_a0}
\end{figure*}

\begin{figure*}
\centering
\includegraphics[width=\linewidth]{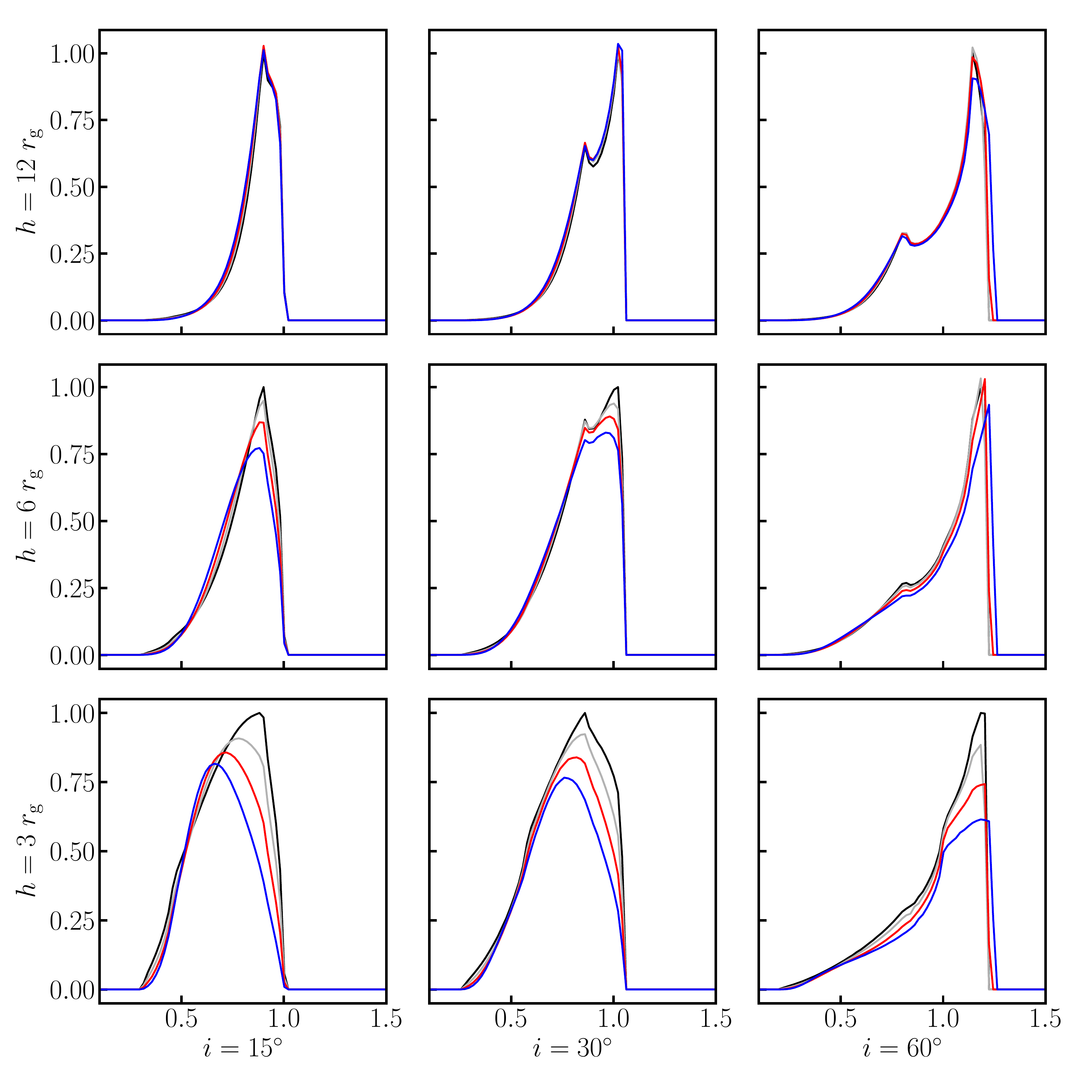}
\caption{A collection of line profiles, similar to Figure \ref{trans_a0}, but for the case of a moderately spinning black hole ($a = 0.9$).  The panels are set up as before, and the color coding represents the value of the Eddington accretion ratio.  For the $i = 15\degree$ and $30\degree$, with $h = 3\,r_{\rm g}$ and $6\,r_{\rm g}$, one sees that there is a shifting of the line centroid towards lower energies with increasing disk thickness, while simultaneously being a slight suppression of the red tail. In the case of $i = 60\degree$, the blue edge of the line appears shift slightly to higher energies for the the low and moderate coronal heights.  Also, in the case of this shallow angle, for $h = 3\,r_{\rm g}$, the blue edge is dampened quite substantially with increasing accretion rate.  In the case of $h = 12\,r_{\rm g}$ for all three values of $i$, one finds that finds only modest differences between the four different disk thicknesses.}
\label{trans_a09}
\end{figure*}

\begin{figure*}
\centering
\includegraphics[width=\linewidth]{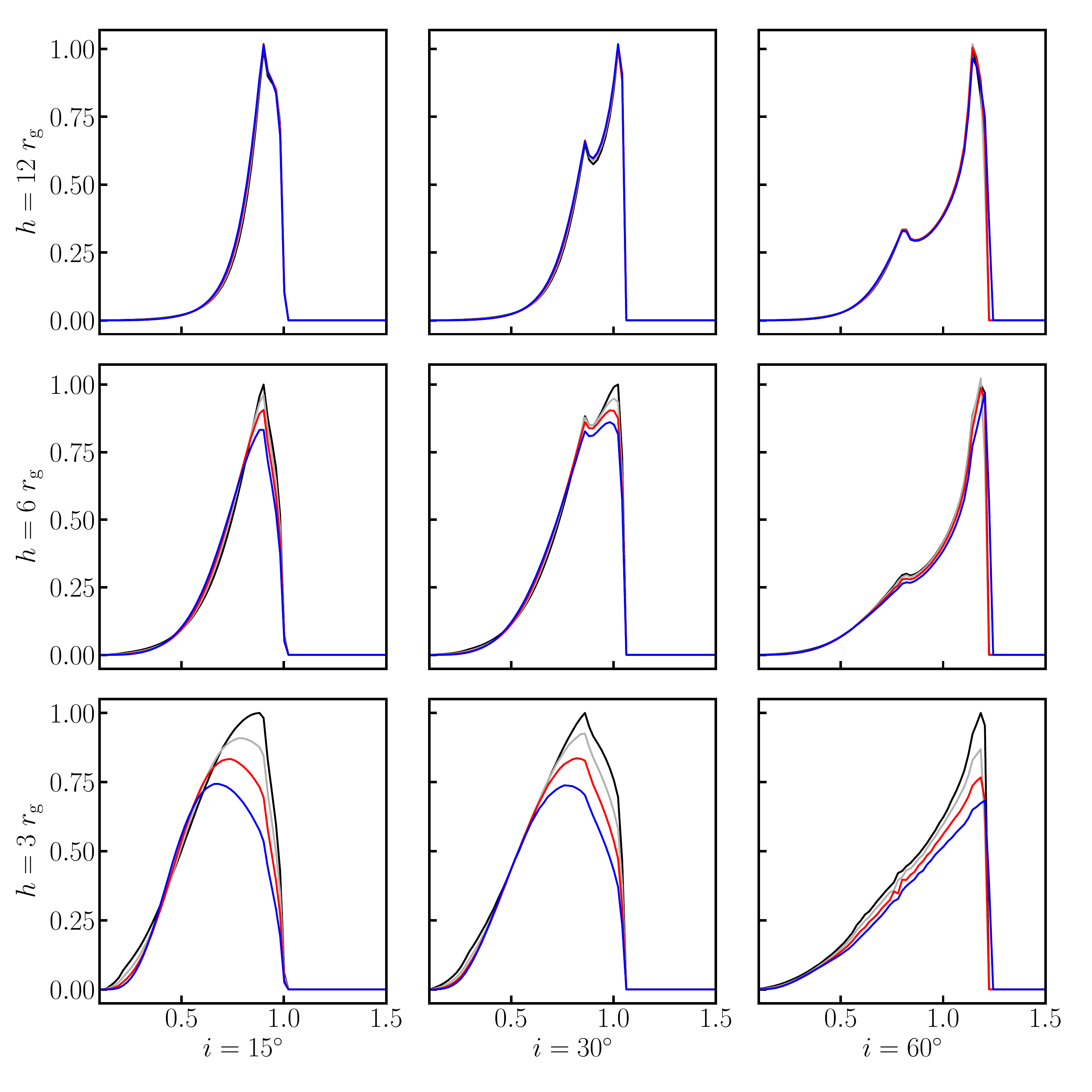}
\caption{Line profiles, similar to Figure \ref{trans_a09}, but for a rapidly spinning black hole ($a = 0.99$).  As in the case of $a = 0.9$ in the mentioned previous figure, one sees an overall shifting of the line centroid towards lower energies for small-to-moderate inclination angles and $h = 3$ or $6\,r_{\rm g}$, while there is an overall dampening of the blue edge for the case of a steeper angle and a corona situated at $h = 3\,r_{\rm g}$.  The overall effects are not as prominent as the moderately spinning black hole due to the decrease in disk thickness from the increase in the efficiency factor ($\eta$) that accompanies the increase in $a$.}
\label{trans_a099}
\end{figure*}

As explained in Section \ref{sec: methods}, one produces the line profile by combining the two separate outputs of our raytracing via equation \ref{eq:lineprofile}.  The line profile can be used as the kernel of the convolution that blurs the rest frame reflection spectrum into that which is seen by the distant observer. Example line profiles are presented in Figures \ref{trans_a0}, \ref{trans_a09}, and \ref{trans_a099} for spin parameters $a =$ 0.00, 0.90, and 0.99 respectively for various values of $h$ and $i$.  The different colors represent the different values of $\dot{M} / \dot{M}_{\rm Edd}$ that we used in the previous sections.  The overall effects of disk thickness on the line profile are quite complex, and is dependent upon the other parameters of the problem.  In the discussion that follows, note that line profiles have been normalized such that the razor-thin disk ($\dot{M} = 0.0$) peaks at unity, allowing one to compare the line intensities across the different disk thicknesses.

For the lowest inclination ($i = 15\degree$) in the Schwarzschild case (Figure \ref{trans_a0} in the left hand side), there are only small changes in the overall line shape.  For a corona at $h = 12\ r_{\rm g}$, the overall of the intensity of the line increases.  For a corona deeper in the potential well ($h = 6\ r_{\rm g}$ and $3\ r_{\rm g}$), one sees a slight shifting of the peak of the line profile by, at most, $\sim$ 10\% towards lower energies. In these cases, one also sees a general change in the line shape, with a suppression of some of the blue edge and an enhancement of the red tail. For $h = 12\ r_{\rm g}$ for moderate and high inclinations ($30\degree$ and $60\degree$ respectively), there is a very apparent enhancement of the high energy peak of the line, with the peaks' luminosity increasing approximately by 25\% and 40\% respectively.  In the latter case, there is a notable shifting of the peak towards higher energies.  For $h = 3\,r_{\rm g}$ and $h = 6\,r_{\rm g}$ for $i = 30\degree$, there is some apparent shifting of the peak, while there also appears to be an enhancement towards the red, the tails developing a more gradual slope than the razor-thin case.  Finally for $i$ = 60\%, one finds both an enhancement towards the blue peak and a softening of the slope in the tail.  Most interesting however, is the relative strength of the luminosity of the blue peak, increasing from $\dot{M} / \dot{M}_{\rm Edd} =$ 0.0 to 0.1, but then decreasing between the $\dot{M} / \dot{M}_{\rm Edd} =$ 0.2 to 0.3 cases.  As observed in the disk images above, there is an apparent balance between the increased intensity due to the change in geometry and the shadowing effects that can now take place once finite thickness is considered. This is a manifestation of this shadowing effect overcoming the increase in intensity, resulting in the suppression of the blue peak.  The fact that this effect is so apparent in the Schwarzschild case is due to the relatively low efficiency at which energy is radiated away.  As per Equation \ref{eq:ss73}, when $\eta$ decreases, the scale height of the disk increases, and thus increasing disk thickness and the ability of the inner regions of the disk to shield the outer regions from the coronal X-rays.

In Figure \ref{trans_a09} we show line profiles for a spin parameter of $a = 0.9$; the effects of disk thickness on the line profile are far more straight-forward than the Schwarzschild case presented in Figure \ref{trans_a0}.  For the case of $h = 12$ r$_{\rm g}$, one observes that the the effects of disk geometry on the line profile are minor for lower inclinations, with some change in the trough in the case of moderate inclinations.  For the case of the high inclination case, one finds a small change in the intensity of the blue peak ($< 10\%$) and a shifting of the blue edge towards higher values of $g$; similar shifting of the blue edge is apparent for the other two values of $h$.  For $h = 6$ r$_{\rm g}$, one finds a suppression of the peak with increasing accretion rate for all inclinations, and a small change of peak wavelength with disk thickness.  These effects are most extreme in the case of $h = 3$ r$_{\rm g}$, with quite dramatic a suppression at high energies ($> 30$\% in the case of high inclination).  For low and moderate inclinations, one sees an overall shifting of the peak energy, with a decrease of approximately 40\% and 30\% respectively.  Finally, for the high inclination, moderate coronal height case,  one finds that the luminosity at the peak increases between $\dot{M} / \dot{M}_{\rm Edd} = 0.0$ and $0.1$, but then decreases with increasing disk thickness, similar to what was found in the right-most column in Figure \ref{trans_a0}

Finally, for a spin of $a = 0.99$ (Figure \ref{trans_a099}), one sees a similar effects as to that of $a = 0.9$, with the geometry effects being minor in the case of a corona with $h = 12$ r$_{\rm g}$ and steadily increase with decreasing $h$.  In the cases with $h = 3 r_{\rm g}$, one finds a general decrease in the line peak intensity, with a suppression of approximately 20\% in the low and moderate inclination cases and a 25\% suppression in the high inclination case.  In the low and moderate inclination cases at this low coronal height, one also sees a decreasing of the line peak energy by approximately 40\% and 10\% respectively.  Similarly, for the  cases of $h = 6 r_{\rm g}$ and $i = 15\degree$ and $30\degree$, one see an overall suppression of the line peaks of 10-15\%.  At this coronal height, in the $i = 60\degree$ case, one sees a similar effect as one sees with these parameters with a $a = 0.9$ black hole (see Figure \ref{trans_a09}), with an initial increase in peak line intensity with disk thickness, which then is suppressed with increasing $\dot{M}$ beyond 0.1 $\dot{M}_{\rm Edd}$.  Also similar to Figure \ref{trans_a09}, one finds that, in the $i = 60\degree$ cases, one observes a slight shift in the blue edge towards higher energies.


Overall, while the effects of disk thickness on the line profile are complex, there appears to be at least a few main effects that are most apparent in the $a = 0.9$ and 0.99 cases.  For low and moderate inclinations and small and moderate coronal heights, there is a suppression of the flux in the blue peak, and a shifting of the peak towards lower energies.  At these high spin values for high inclinations, one finds an overall shift in the blue edge towards higher energies.  These two cases are the most relevant for the study of supermassive black holes at the heart of well-studied AGN; the observed population of AGN is dominated by high-spin objects \citep{Reynolds2014} understood as a result of the natural efficiency bias in flux limited surveys \citep{Brenneman+2011,Vasudevan+2016}.

\subsection{Testing For Bias} \label{sec:delta-chi}

In the previous section, we have examined the effects of disk thickness on the shape of the line profile, demonstrating that the blurring kernels created using a finite-thickness disk model are notably different than those generated while approximating the disk as razor thin. Assuming our model accurately describes a physical system (e.g. the AGN of a Seyfert I galaxy), disk thickness effects would imprint themselves on the observed reflection spectra, exhibiting features (e.g. a broad Fe K$\alpha$ fluorescence line) that are qualitatively different from past predictions.  This could potentially result in bias in attempts to estimate parameters (e.g. the spin of the black hole) from fitting previous models to real data, and in this section we will make a short assessment of this issue. We have chosen to focus on a specific test case to illustrate the importance of disk geometry when estimating black hole spin using reflection modeling in a fairly typical AGN; an exhaustive exploration of parameter space is outside the scope of this particular study. For this analysis, we focused on the cases of $\dot{M}$/$\dot{M}_{\rm Edd}$ $\in \{0.0,\ 0.1,\ 0.2,\ 0.3\}$.

We imported line profiles generated by {\tt Fenrir} into {\tt XSPEC} \citep{Arnaud1996} by making use of a slightly modified version of {\tt relconv\_lp}, a convolution lamp post model available in the {\tt RELXILL} package \citep{Garcia+2014,Dauser+2014}.  We modified the {\tt RELXILL} source code to insert our own line profiles as the convolution kernel, approximating the spectroscopic X-ray features of an Seyfert 1 AGN by convolving it with the atomic disk model {\tt xillver} \citep{Garcia+Kallman2010,Garcia+2011,Garcia+2013}, combining the result with a power-law (spectral index of $\Gamma = 2$) and a standard absorption model ({\tt phabs}, $N_{\rm H} = 5\cdot10^{20}\ {\rm cm}^{-2}$). The chosen model parameters can be found in Table \ref{params}, where the models were normalized so that the total flux between 2-10 keV is $\sim 6\times10^{-11}$ erg cm$^{-2}$ s$^{-1}$.  From these models, we created mock data sets, simulating 200 ks observations with XMM (2-10 keV) using {\tt XSPEC}, which were binned to such that  the number of photons-per-bin $> 20$ to better allow the use of $\chi^{2}$ statistics.

\begin{table}
\centering
\caption{Model Parameters Used For Data Creation And Analysis}
\begin{tabular}{ c | c | c || c | c | c}
Model & Parameter & Value & Model & Parameter & Value \\
\hline
\hline
{\tt phabs} & $N_{\rm H}$ & $\lbrack 5\cdot10^{20}\ \rm{cm}^{-2} \rbrack$ & & $\rho_{\rm out}$ & $\lbrack 100\ r_{\rm g} \rbrack$ \\
{\tt powerlaw} & $\Gamma$ & 2 & {\tt xillver} & $A_{\rm Fe}$ & 3.7 \\
& Norm & $2\cdot10^{-2}$ photons keV$^{-1}$ cm$^{-2}$ s$^{-1}$ & & $E_{\rm cut}$ & $\lbrack 300\ \rm{keV}\rbrack$  \\
{\tt Fenrir} & h & 3 $r_{\rm g}$ & & $log(\xi)$ & 1.0 \\
& a & 0.9 & & z & $\lbrack 0 \rbrack$ \\
& i & $15\degree$ & & $R_{\rm f}$ & $\lbrack -1 \rbrack$ \\
& $\dot{M}$/$\dot{M}_{\rm Edd}$ & 0.0, 0.1, 0.2, 0.3 & &Norm  & $5.5\cdot10^{-4}$ photons keV$^{-1}$ cm$^{-2}$ s$^{-1}$\\
& $\rho_{\rm in}$ & $\lbrack r_{\rm ISCO} \rbrack$ & & \\
\hline
\end{tabular}
\vspace*{2mm}\par{}
{\parbox{1.0\textwidth}{\footnotesize{}The values of the model parameters used to create our set of mock XMM data. We have included a simple absorption model ({\tt phabs}), a power-law model ({\tt powerlaw}), and a model of the reflection spectrum as seen in a frame co-moving with the disk ({\tt xillver}). The spectrum from {\tt xillver} is convolved with a blurring kernel generated by {\tt Fenrir}, imported into XSPEC using a modified version of {\tt relconv\_lp}, a part of the {\tt RELXILL} package. Square brackets correspond to values that are held constant throughout the analysis. The normalization parameters were chosen to represent a fairly typical Seyfert-I galaxy with a total flux of $\sim 6\times10^{-11}$ erg cm$^{-2}$ s$^{-1}$ between 2-10 keV.}}
\label{params}
\end{table}

\begin{figure*}
\centering
\includegraphics[width=0.48\linewidth]{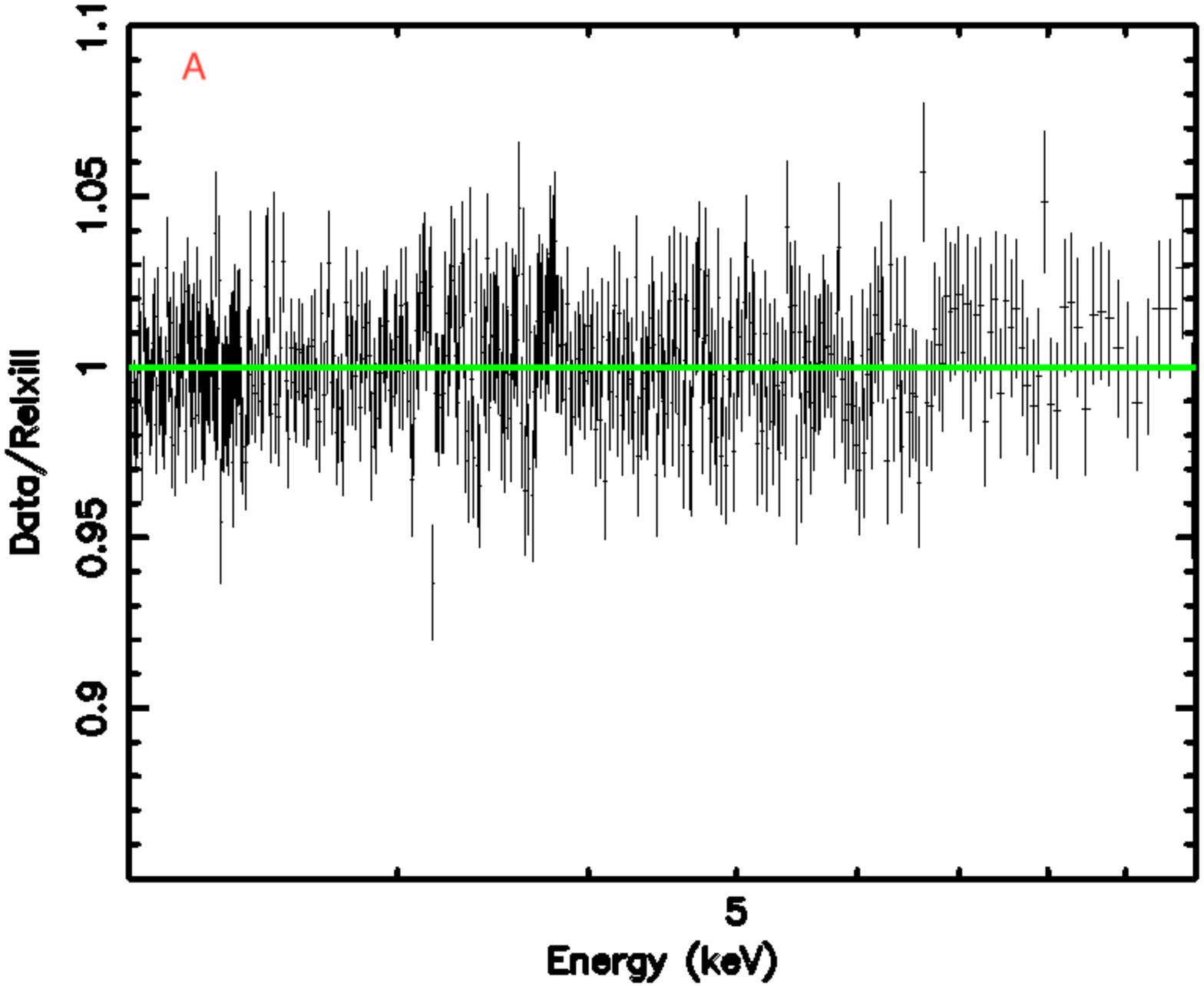}
\includegraphics[width=0.48\linewidth]{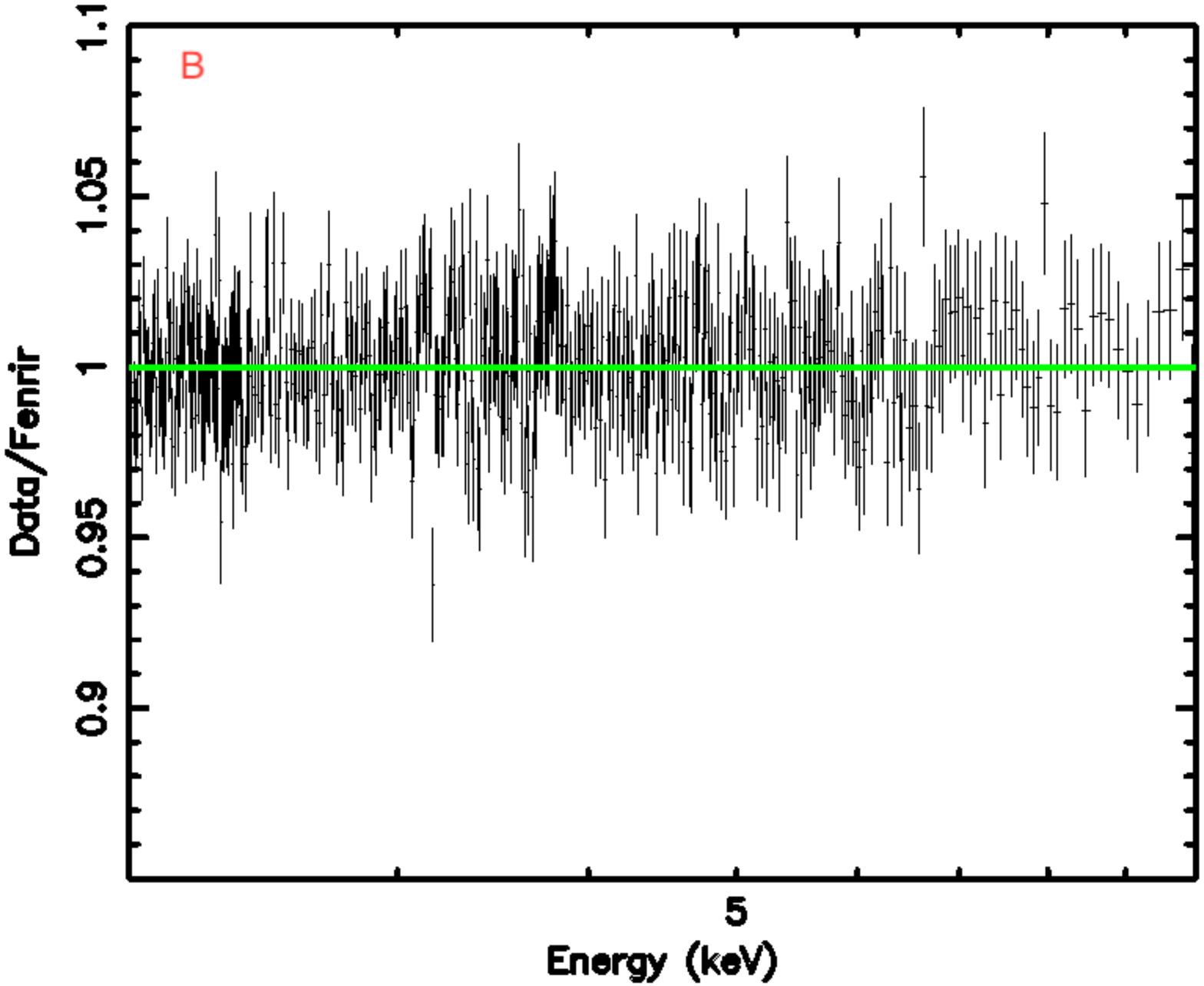}
\includegraphics[width=0.48\linewidth]{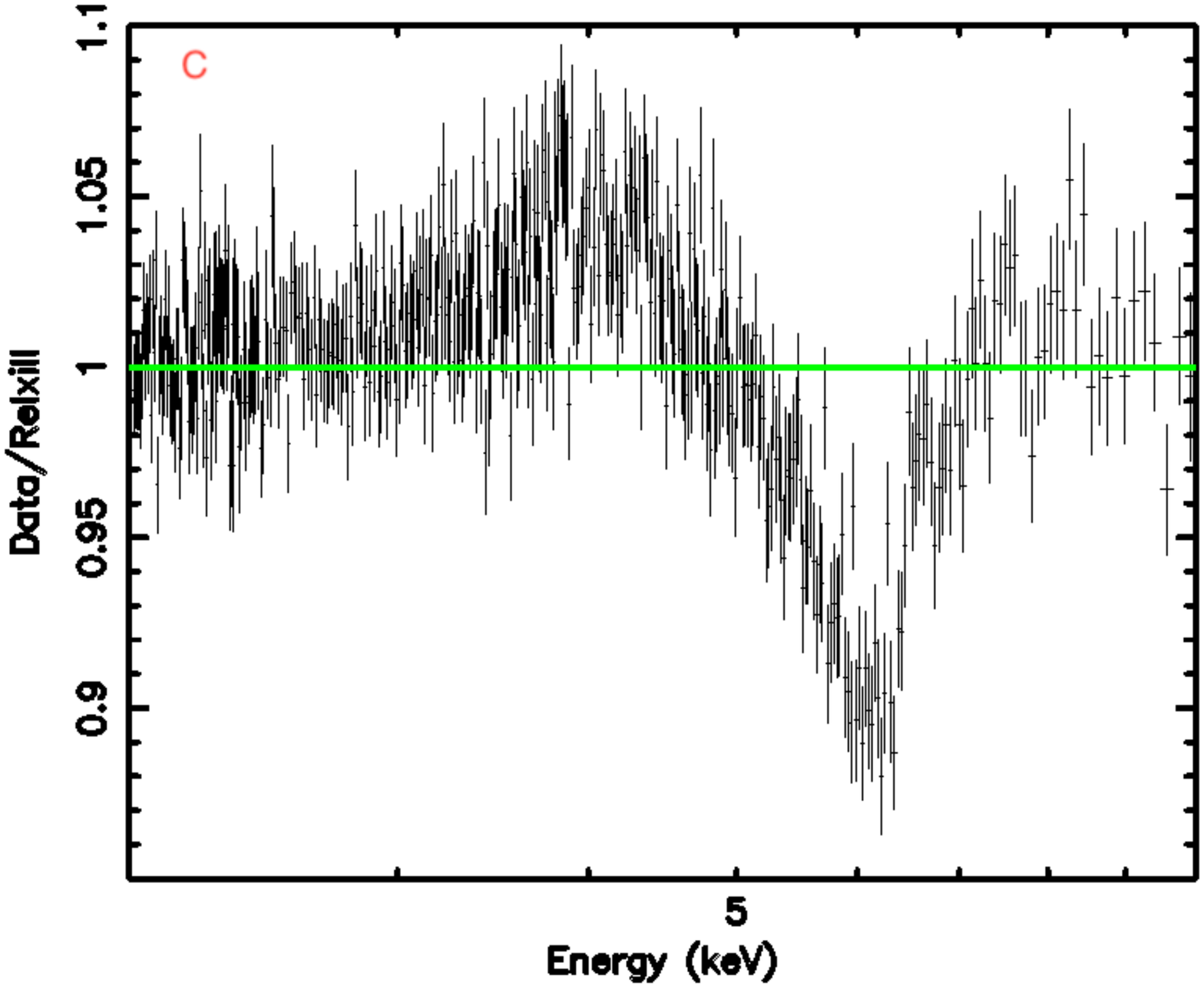}
\includegraphics[width=0.48\linewidth]{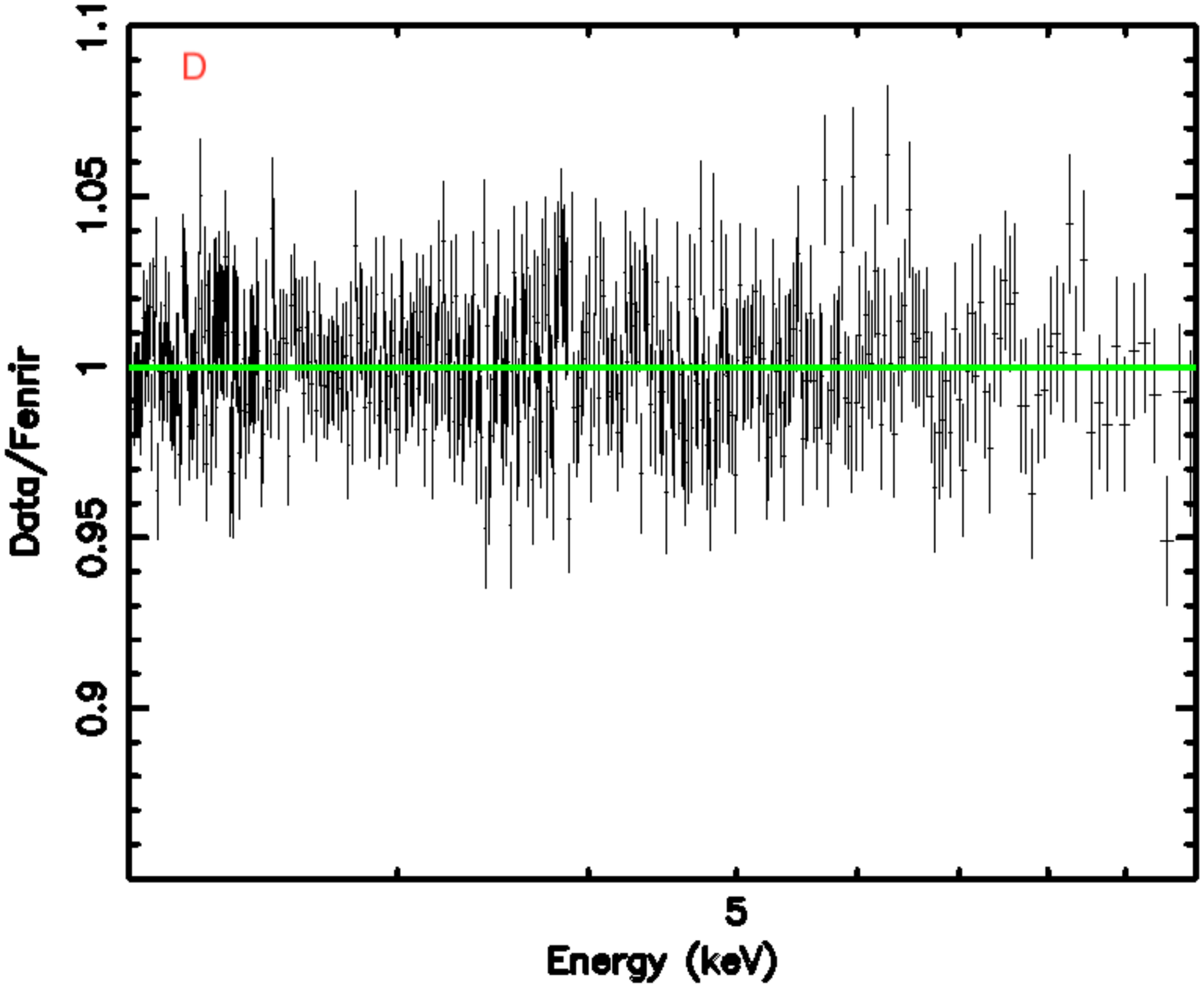}
\caption{Ratios between 200 ks of simulated AGN X-ray spectral data [created using the model in Table \ref{params} using $\dot{M}$ = 0.0 $\dot{M}_{\rm Edd}$ (top row) and 0.3 $\dot{M}_{\rm Edd}$ (bottom row)] and models from {\tt RELXILL} (left column) and {\tt Fenrir} (right column) that use the same input parameters. The data is generated using XMM (2-10 keV) response matrices, and the corresponding $\chi^2$ with each case is given in Table \ref{chi2vals}.  As one can see from panels A and B, both {\tt RELXILL} and {\tt Fenrir} are qualitatively consistent with the case of a $\dot{M}$ = 0.0 $\dot{M}_{\rm Edd}$, which is mathematically equivalent to a razor-thin disk. However, in the case of $\dot{M}$ = 0.3 $\dot{M}_{\rm Edd}$ in panels C and D, {\tt RELXILL} is insufficient in accurately modeling the X-ray spectra.  This emphasizes the importance of using finite-thickness disk models (like {\tt Fenrir}) in the spectral analysis of AGN.}
\label{ratios}
\end{figure*}

After creating this simulated data in {\tt XSPEC} for each of the accretion rates of interest, we compared each data set with razor-thin disk models from (unmodified) {\tt RELXILL} using the same input parameters listed in Table \ref{params}.  Likewise, we compared this data to the corresponding model from {\tt Fenrir} as a control, using the value of $\dot{M}$ used to generate the particular data set. Figure \ref{ratios} illustrates the case of an razor-thin disk ($\dot{M}$ = 0.0 $\dot{M}_{\rm Edd}$) and a disk with a corresponding $\dot{M}$ = 0.3 $\dot{M}_{\rm Edd}$, showing that while {\tt RELXILL} is visually consistent with the data in the former case, the {\tt RELXILL} predictions overestimate the flux near $\sim 6$ keV while underestimating the flux $< 5$ keV in the latter. This is consistent with the reddening of the convolution kernel shown in Figure \ref{trans_a09}. As expected, {\tt Fenrir} is consistent with the data in both cases.

For each value of $\dot{M}$ we calculated the corresponding $\chi^2$ between the data and the two models, allowing for a $\Delta \chi^2$ comparison between the {\tt RELXILL} ($\chi^{2}_{\rm r}$) and {\tt Fenrir} ($\chi^{2}_{\rm f}$). The results of this analysis is presented in Table \ref{chi2vals}, where the degrees of freedom in each case (dof) is equal to the difference between the number of data bins and the free parameters of the model.  As suggested by the qualitative analysis in Figure \ref{ratios}, both {\tt RELXILL} and {\tt Fenrir} are statistically consistent with the simulated data and with each other when $\dot{M} = 0.0\ \dot{M}_{\rm Edd}$, however {\tt RELXILL} is inconsistent with the three other data sets  where $\dot{M} \neq \ 0.0 \dot{M}_{\rm Edd}$.  $\chi^{2}_{\rm r}$ is increasingly worse with increasing $\dot{M}$, and the $\Delta \chi^{2}$ values likewise show that {\tt RELXILL} and {\tt Fenrir} have divergent predictions for systems where disk thickness is non-negligible.


\begin{table}
\centering
\caption{$\Delta \chi^{2}$ Comparison Between {\tt Fenrir} And {\tt RELXILL}}
\begin{tabular}{c || c | c | c | c || c | c}
$\dot{M}$/$\dot{M}_{\rm Edd}$ & $\chi^{2}_{\rm r}$ & dof$_{\rm r}$ & $\chi^{2}_{\rm f}$ & dof$_{\rm f}$ & $\Delta \chi^{2}$ & $P$ \\
\hline
\hline
0.0 & 1472.95 & 1591 & 1472.93 & 1590 & 0.02 & 0.8875 \\
0.1 & 1707.88 & 1591 & 1633.14 & 1590 & 74.74 & $\sim 0$ \\
0.2 & 1991.21 & 1591 & 1630.44 & 1590 & 360.77 & $\sim 0$ \\
0.3 & 2707.58 & 1591 & 1582.55 & 1590 & 1125.03 & $\sim 0$ \\
\hline
\end{tabular}
\vspace{2mm}\par{}
{\parbox{1.0\textwidth}{\footnotesize{}Comparison between 200 ks of XMM data generated from {\tt Fenrir} and models from {\tt RELXILL} and {\tt Fenrir} with the same parameter values used to generate the data (see Table \ref{params}). Each data set is characterized by an Eddington accretion ratio (left-most column), where the disk thickness increases with increasing accretion rate. The $\chi^{2}$ values ($\chi^{2}_{\rm r}$ for {\tt RELXILL} and $\chi^{2}_{\rm f}$ for {\tt Fenrir}) show that both models can describe the case of $\dot{M}$ = 0 $\dot{M}_{\rm Edd}$) (equivalent to a zero-thickness disk) sufficiently well. In the case of a finite thickness disk, {\tt RELXILL} clearly is not sufficient in describing the simulated data, with the $\Delta \chi^{2}$ values being significant in all three non-zero accretion rates ($P$ values in the right-most column). Together, this comparison shows that fitting a model that neglects disk thickness (such as {\tt RELXILL}) to real data would result in significant bias.}}
\label{chi2vals}
\end{table}

Taken together with Sec \ref{sec:disks} and \ref{sec:lines}, one can conclusively say that disk geometry (and disk thickness in particular) can have observable effect on the features of the reflection spectrum and, if not properly accounted for, could significantly bias any attempt to estimate the physical parameters of AGN. A simple $\Delta \chi^{2}$ analysis shows that, if AGN have appreciable disk thicknesses, using a model that approximates the disk as being razor-thin (e.g. {\tt RELXILL}) is likely to be inaccurate once the Eddington ratio exceeds 0.1.

\begin{figure*}
\centering
\includegraphics[width=0.96\linewidth]{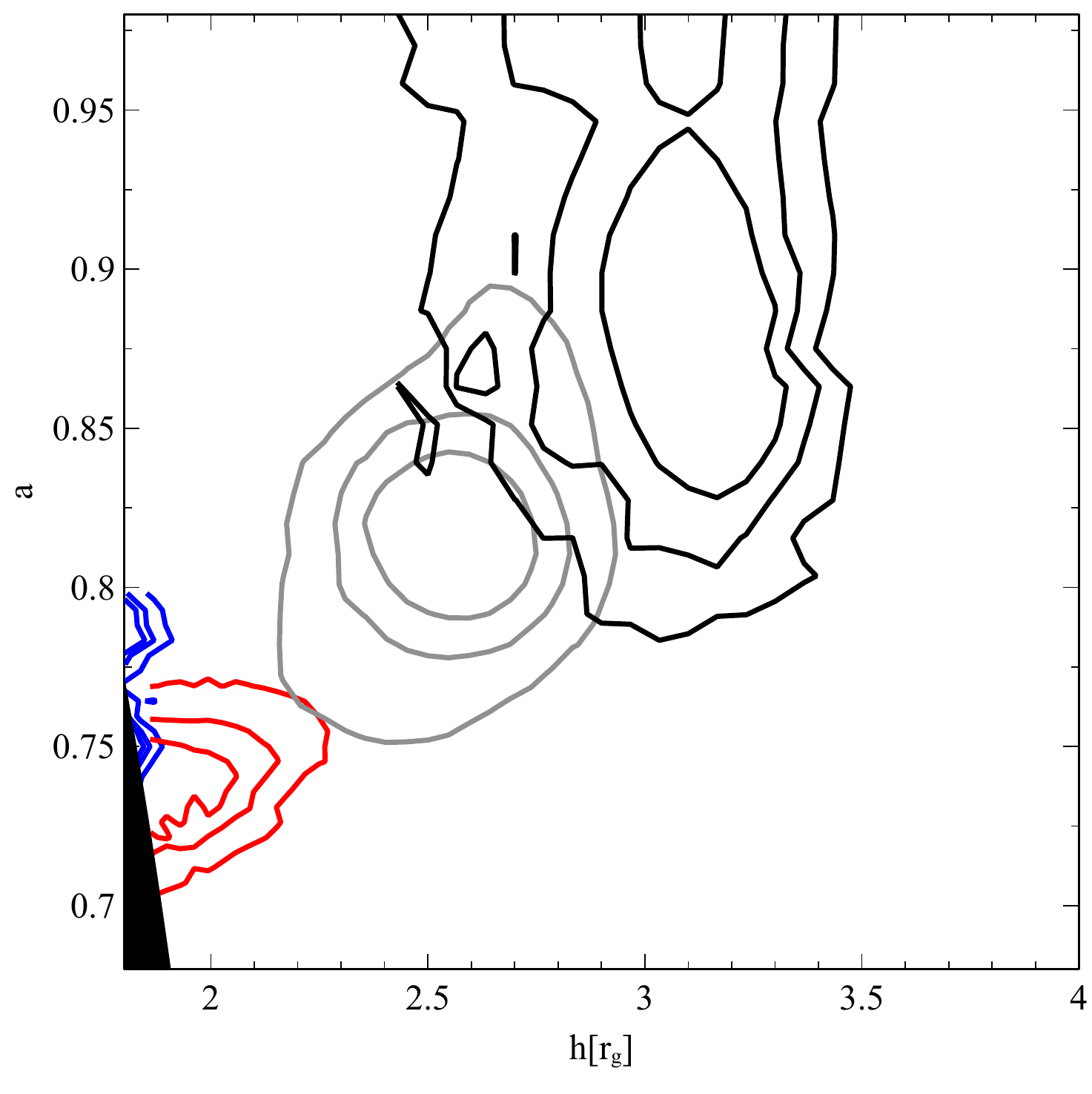}
\caption{$\chi^{2}$ contours of coronal height ($h$) compared black hole spin ($a$), with the levels showing the 68\%, 90\%, and 99\% confidence regions. We created the data by simulating 200 ks XMM observations using {\tt Fenrir} with the model parameters listed in Table \ref{params} and is fit using {\tt RELXILL} and using these same parameters as initial values of the fit. As previously, each color represents a different different disk thickness: zero-thickness ($\dot{M}$ = 0.0 $\dot{M}_{\rm Edd}$, black), $\dot{M}$ = 0.1 $\dot{M}_{\rm Edd}$ (gray), $\dot{M}$ = 0.2 $\dot{M}_{\rm Edd}$ (red), and $\dot{M}$ = 0.3 $\dot{M}_{\rm Edd}$ (blue). Finally, the solid black region in the lower left represents the lower limit of $h$ in {\tt RELXILL} (1.1$r_{\rm +}$, where $r_{\rm +}$ is the radius of the event horizon). While we were unable to find a statistically good fit for the $\dot{M}$ = 0.3 $\dot{M}_{\rm Edd}$, we included this case for completeness. The cases of $\dot{M}$ = 0.1 $\dot{M}_{\rm Edd}$ and 0.2 $\dot{M}_{\rm Edd}$ conclusively show that, for this region of parameter space, neglecting the thickness of the disk results in a significant underestimate of both $a$ and $h$, even in the case where one knows the true parameters of the system a priori.}
\label{contours}
\end{figure*}

To further explore these systematic errors, we fit each set of simulated data using {\tt RELXILL}, assuming we knew the true parameters a priori and using said parameter values as our initial values of the fit (see Table \ref{params}). Using standard {\tt XSPEC} spectral fitting routines, we were successful in fitting both $\dot{M}$ = 0.0 $\dot{M}_{\rm Edd}$ and 0.1 $\dot{M}_{\rm Edd}$ (P $\sim$ 0.99 and 0.24 respectively), and $\dot{M}$ = 0.2 $\dot{M}_{\rm Edd}$ had a modest fit (P $\sim$ 0.03). The most extreme case of $\dot{M}$ = 0.3 $\dot{M}_{\rm Edd}$, however, proved difficult to fit successfully with {\tt RELXILL} (P $\sim$ $10^{-8}$). We present $\chi^{2}$ contours for black hole spin and coronal height in Figure \ref{contours}, giving the 68\%, 90\%, and 99\% confidence intervals for each value of $\dot{M}$ (each represented by different colors). A quick examination of the figure shows that, as the accretion rate increases and thus the disk becomes thicker, both $a$ and $h$ become increasingly underestimated, with the razor-thin disk ($\dot{M}$ = 0.0 $\dot{M}_{\rm Edd}$, black) and $\dot{M}$ = 0.1 $\dot{M}_{\rm Edd}$ (gray) having a $\Delta a$ $\sim$ 0.1 and $\Delta h$ = 0.5 $r_{\rm g}$. The $\dot{M}$ = 0.2 $\dot{M}_{\rm Edd}$ case (red) exhibits contours that suggest a corona that hugs the event horizon, the contours confined in the lower-left region of the plot near the lower limit of $h$ = 1.1 $r_{\rm +}$ (represented by the solid black region in the lower-left corner); $r_{\rm +}$ is the event horizon of the black hole which decreases with increasing $a$. The contours for $\dot{M}$ = 0.3 $\dot{M}_{\rm Edd}$ are shown for completeness.

We do not claim that this method is a rigorous recreation of observational spectral fitting, as we would naturally not know what the true values prior to the analysis, or at the very least we would have some uncertainty in our prior even if there was previous independent measurements. Instead, this represents an ideal condition, giving {\tt RELXILL} the best chance of reproducing the true parameter values and disentangling the systematic errors of interest from those inherent in the choice of initial parameters in a finite exploration of multi-dimensional parameter space. We have a conclusive proof of principle, and we will perform a more rigorous exploration of systematic errors using {\tt Fenrir} in the near future.

\section{Discussion} \label{sec:discussion}

While our investigation has relied on a likely overly-simplistic model of disk structure, we have a proof of principle that the geometries of black hole accretion disks likely have significant effects on observed reflection spectrum. It is due to these effects that the use of the razor-thin disk approximation has likely created systematic errors in currently reported values of parameters of astrophysical interest, and a thorough investigation of this is warranted. In the following we section, we discuss how non-trivial disk geometry may have already been seen, giving examples of observations that may be more easily explained by {\tt Fenrir} than by a model that neglects the vertical extent of the accretion disk. While the following discussion in no way claims to be exhaustive, we hope to further illustrate the importance in venturing beyond razor-thin disks.

\cite{Kara+2015} presents observations of the Seyfert I galaxy, 1H0707-495, which has been shown to have the curious feature of an Fe K$\alpha$ with a blue peak whose energy correlates with X-ray luminosity ($L_{\rm X}$) in low X-ray states (flux in the 3-10 keV band of $< 8 \times 10^{-13} \,erg\,s^{-1}\,cm^{-2}$). One possible explanation of this is a change of inclination of the disk as a function of time, with a less-inclined disk having a naturally less energetic blue edge due to the reduced Doppler shifts.  This scenario is unlikely however, as it would require a change of $\Delta i > 10\degree$ on the scale of a few days.  Other possible explanations have been proposed as to the origin of this phenomena, including the changing of a partial covering absorber \citep{Gallo+2004} and a dramatic change in the emissivity profile \citep{Fabian+2004}.  \cite{Kara+2015} argued for the preference of the latter over the former, suggesting that a change in coronal height or a non-trivial disk geometry could explain this phenomenon. The results in the previous sections do give credence to this hypothesis, as we have shown that, even with a simple disk geometry, a change in coronal height (which would naturally modulate $L_{\rm X}$) can result in the shifting of the peak (see Figures \ref{trans_a0}-\ref{trans_a099}.  Such a shift over time could be interpreted as a wobble in the disk inclination if one operates with the assumption of a razor-thin disk, while in reality it could easily be explained by internal variations within the corona.

Another possible systematic error that could be inherent in the use of previous models is an underestimation of the disk plasma ionization parameter ($\xi$).  \cite{Ballantyne+2011} presented an analysis of the ionization states for a moderate-sized sample of AGN that have both a rigorous mass measurement and clear signs of reflection, showing that power-law relationship between $\xi$ and the Eddington luminosity ratio ($L_{\rm bol}/L_{\rm Edd}$) is shallower than would be expected from classical plasma and accretion physics. While there are a number of hypotheses as to the nature of this discrepancy between the inferred and predicted relationships discussed by the authors, one possible explanation could be that the reddening of the line profile at increasing disk thicknesses (i.e. at increasing Eddington ratios) could result in a shallowing of the Eddington ratio-ionization relation from that which is predicted by theory. As discussed by \cite{Ballantyne+2011}, $\xi$ is inferred by the blue peak of the broad Fe line, which shifts to higher energies with increasing levels of ionization.  If one has a reddening of the line profile with increasing disk thickness, where much of the blue intensity is suppressed and there is enhancement towards lower energies, you would naturally underestimate the value of $\xi$.

A past study that examined the effects of disk structure on the reflection spectrum was performed by \cite{Pariev+Bromley1998}, the authors focusing on the shape and intensity of the Fe K$\alpha$ fluorescence line from a fully relativistic accretion disk \citep{Novikov+Thorne1973}. They assumed the photosphere is defined by the surface at which the density of the disk drops to zero, and included both turbulence and a net radial velocity of the disk material ($u^{r}\ \neq 0$). However, unlike our analysis which approximated the corona as a lamp-post, the authors assume that the emissivity of the disk is a simple power-law model ($\propto \rho^{-q}$) where $q \in [2,4]$. These differences in the treatment of the emissivity profile results in qualitatively different effects: the prevalent geometric affect observed by the authors was the overall decrease in the flux of the Fe K$\alpha$ line near the blue peak in the cases of extreme Kerr ($a = 0.998$) and a Schwarzschild ($a = 0.0$) black holes. This seems inconsistent with what we found in Figure \ref{trans_a0}, where non-zero disk thickness was shown to result in a much more luminous blue peak in certain instances. The work also noted an overall reddening and broadening of the line due to the radial and turbulent motions of the gas, but the former effect is far less dramatic than observed using {\tt Fenrir} (which neglects bulk radial motions).

A later work by \cite{Wu+Wang2007} approximated the disk cross-section as triangular in shape, defined by some opening angle, and a modified velocity field to mimic sub-Keplerian orbits at high accretion rates. Like \cite{Pariev+Bromley1998}, they assume a power-law emissivity profile $q$ = 3, but they included various limb darkening prescriptions that have substantial effect on the predicted line profiles. For a spin of $a$ = 0.998 at $i$ = $30\degree$, the authors find that increasing the disk thickness shifts the location of the blue peak towards lower energies, qualitatively consistent with what we found in Figure \ref{trans_a099}. They also present disk images that likewise have the "bowl" geometry exhibited by our own disk images in Figure \ref{gDisks}, with the disk eventually eclipsing the black hole and the inner-most radii fo the disk.

All together, comparing our work with that of \cite{Pariev+Bromley1998} and \cite{Wu+Wang2007} suggests that the predicted signatures that disk geometry imparts on the reflection spectra is dependent how one models the corona. The most obvious example is that of disk self-shielding (see Figure \ref{iDisks}), where the inner regions of the disk act to shield the outer regions from coronal X-rays. If one were to use a power-law emissivity profile (e.g. $\mathcal{E} \propto \rho^{-3}$) or calculate $\mathcal{E}$ from corona that is a highly-ionized disk atmosphere, this effect would not be present. Thus, if one were to accurately model the structure of an AGN accretion disk, it is plausible that one could probe the geometry of the corona at an unprecedented level.

\section{Summary} \label{sec:summary}

Responsible for the energetic processes found in AGN, black holes are simple objects that can be described by two fundamental parameters: their mass ($M$) and their spin parameter ($a$). Measuring the spin of the supermassive black holes allows us to understand their underlying physics and their accretion histories. The X-ray reflection spectrum offers a way to measure $a$, as well as to probe the inner-most regions of the accretion disk and the hot electron corona.  Despite the maturity of the field, the model of the accretion disk used in modeling the reflection spectrum has not progressed much beyond that of an optically thick plane of material orbiting in circular Keplerian orbits with no vertical extent. As such we have created a new raytracing suite known as {\tt Fenrir} which allows the user to specify an analytic disk thickness ($z$), using $z$ as a stopping condition for the integration of photon trajectories through Kerr spacetime. In this initial study, we approximated the photosphere as equal to twice the pressure scale height of a radiation pressure dominated, geometrically thin, optically thick \cite{Shakura+Sunyaev1973} accretion disk. Our aim is to examine the effects that disk thickness may have on the reflection spectrum (e.g. the shape of the Fe K$\alpha$ line) and to check for the existence of systemic bias inherent in the use of the razor-thin disk approximation. In this analysis, we approximate the irradiating corona as a lamppost point source at height $h$ along the rotation axis of the black hole that is emitting isotropically in its rest frame.

Examining the irradiation profiles produced by {\tt Fenrir} for various disk thicknesses (labeled by accretion rate, $\dot{M}$/$\dot{M}_{\rm Edd}$ $\in \{0.0,\ 0.1,\ 0.2,\ 0.3\}$), one finds that the inner regions of the accretion disk shield the outer regions from the coronal X-rays, resulting with an enhancement of incident flux at small-to-moderate cylindrical radii while simultaneously reducing the incident flux larger radii. This effect is most prominent when $h$ is small and $a = 0.0$ where the disk is thickest for a given $\dot{M}$, however it is also seen when $a = 0.9$ and $a = 0.99$. When examining reflection intensity profiles mapped over the disk image plane, this self-shielding translates to a change in the size of the emitting region. We also find that there is an overall flattening of the incident flux near $r_{\rm ISCO}$ when $a = 0.9$ and $a = 0.99$ due to convex geometry of the disk.

We show that the line profiles (convolution kernels) likewise are affected by the inclusion of disk thickness, with the shifting of the blue peak towards lower energies when $a = 0.9$, $0.99$; $i = 15\degree$, $30\degree$; and $h = 3 r_{\rm g}$, $6 r_{\rm g}$, with the most dramatic being the case of $a = 0.9$, $i = 15\degree$, and $h = 3 r_{\rm g}$ with the peak shifting by $\sim 40\%$. There was also an increase in the blue peak intensity at $i = 60\degree$ for all values of $a$ (save for $a = 0.99$ and $h = 3 r_{\rm g}$ which decreased in flux with increasing $\dot{M}$), as well as a slight shifting of the peak to edge towards higher energies. This second effect is not a direct correlation between the flux at the blue peak and disk thickness however, with the line being suppressed beyond a certain $\dot{M}$. When $h = 12 r_{\rm g}$, there were only modest changes to the line profile.

We simulated four sets of 200 ks of XMM data (2-10 keV) using {\tt Fenrir} at the various levels of $\dot{M}$ of interest and performed a $\Delta \chi^{2}$ analysis of the commonly used {\tt RELXILL} model (which makes use of the zero-thickness disk approximation) and {\tt Fenrir}. Overall, we find that while both models do well at modeling the case where $\dot{M} = 0.0\ \dot{M}_{\rm Edd}$ (which reduces down to the zero-thickness approximation), {\tt RELXILL} performs progressively worse at modeling data as disk thickness increases. The $\Delta \chi^{2}$ between {\tt RELXILL} and {\tt Fenrir} is significant in all three cases where $\dot{M} \neq 0.0\ \dot{M}_{\rm Edd}$. This suggests that using {\tt RELXILL} to model data taken from a system with non-negligible disk thickness would result in significant biases. As a proof of principle, we fit each data set with {\tt RELXILL}, assuming the ideal scenario that we knew the true characteristics of the system a priori. Examining the $\chi^{2}$ contours for black hole spin and coronal height, we find that both are underestimated in the region of parameter space being examined. A more thorough and rigorous exploration of systematic errors is warranted, and is being planned for the near future.
\newline
\newline
We would like to thank Cole Miller, Erin Kara, Drew Hogg, Dom Walton, Matt Middleton, Thomas Dauser, Javier Garc\'ia, Michal Dovciak, Michal Bursa, Dan Wilkins, and Jeremy Sanders for many stimulating and fruitful conversations. We gratefully acknowledge support from NASA under grants NNX17AF29G and NNX15AU54G. Finally, we would like to thank the Department of Astronomy as the University of Maryland for allowing us to use part of their computational resources on the {\tt Deepthought} and {\tt yorp} clusters.

\software{XSPEC \citep{Arnaud1996}, RELXILL \citep{Garcia+2014,Dauser+2014}, ky \citep{Dovciak+2004}, Fenrir}




\end{document}